\documentclass[useAMS,usenatbib,usegraphicx]{mn2e}
\newcommand{\msun}{{\rm M}_{\solar}}

\newcommand{\solar}{\ifmmode_{\mathord\odot}\;\else$_{\mathord\odot}\;$\fi}

\def\ltsima{$\; \buildrel < \over \sim \;$}
\def\lsim{\lower.5ex\hbox{\ltsima}}
\def\Msunh{\mbox{$h^{-1}$M$_\odot$}}
\def\mpch{\mbox{$h^{-1}$Mpc}}

\def\Mpc{{\rm Mpc}}

\def\deg{\ifmmode{^\circ}\else{$^\circ$}\fi}
\def\hGpc{\ifmmode{h^{-1}{\rm Gpc}}\else{$h^{-1}{\rm Gpc}$}\fi}
\def\hkpc{\ifmmode{h^{-1}{\rm kpc}}\else{$h^{-1}{\rm kpc}$}\fi}
\def\hMpc{\ifmmode{h^{-1}{\rm Mpc}}\else{$h^{-1}{\rm Mpc}$}\fi}
\def\hMsun{\ifmmode{h^{-1}M_\odot}\else{$h^{-1}M_\odot$}\fi}

\def\muK{\ifmmode{\mu{\rm{K}}}\else{$\mu$K}\fi}
\def\mum{\ifmmode{\mu{\rm{m}}}\else{$\mu$m}\fi}

\def\sige{\mbox{$\sigma_8$}}

\newcommand{\LCDM}{$\Lambda$CDM}
\newcommand{\aj}{{AJ}}
\newcommand{\apj}{{ApJ}}
\newcommand{\apjl}{{ApJ}}
\newcommand{\apjs}{{ApJS}}
\newcommand{\mnras}{{MNRAS}}

\newcommand{\pra}{Phys. Rev. A}

\title{Statistics of Voids in the 2dF Galaxy Redshift Survey}

\author[Patiri et al.]{
\parbox[t]{\textwidth}{
Santiago G. Patiri$^{1}$\thanks{E-mail:
spatiri@iac.es},
Juan E. Betancort-Rijo$^{1,2}$,
Francisco Prada$^{3}$, 
Anatoly Klypin$^{4}$ \&
Stefan Gottl\"ober$^{5}$
}
\\
\\
$^1$
Instituto de Astrofisica de Canarias,
C/ Via Lactea s/n, Tenerife, E38200, Spain
\\
$^2$
Facultad de Fisica, Universidad de La Laguna,
Astrofisico Francisco Sanchez, s/n, La Laguna
Tenerife, E38200, Spain
\\
$^3$
Ramon y Cajal Fellow, Instituto de Astrofisica de Andalucia (CSIC), E-18008, Granada, Spain
\\
$^4$
Astronomy Department,
New Mexico State University, Dept.\ 4500,
Las Cruces, NM 88003, USA
\\
$^5$
Astrophysikalisches Institut Potsdam,
An der Sternwarte 16,
14482 Potsdam, Germany.
\\
}

\begin{document}

\maketitle

\begin{abstract}
We present a statistical analysis of voids in the 2dF galaxy
redshift survey (2dFGRS). In order to detect the voids, we have
developed two robust algorithms.  We define voids as non-overlapping
maximal spheres empty of halos or galaxies with mass or luminosity
above a given one. We search for voids in cosmological $N$-Body
simulations to test the performance of our void finders. We obtain and
analyze the void statistics for several volume-limited samples for the
North Galactic Strip (NGP) and the South Galactic Strip (SGP)
constructed from the 2dFGRS full data release.  We find that the
results obtained from the NGP and the SGP are statistically
compatible. From the results of several statistical tests we conclude
that voids are essentially uncorrelated, with at most a mild
anticorrelation and that at the 99.5\% confidence level there is a 
dependence of the void number density on redshift. We 
develop a technique to correct the distortion caused by the fact that
we use the redshift as the radial coordinate. We calibrate this
technique with mock catalogues and find that the correction might be
of some relevance to carry out accurate inferences from void
statistics.  We study the statistics of the galaxies inside nine
nearby voids. We find that galaxies in voids are not randomly distributed:
they form structures like filaments. We also obtain the galaxy number density
profile in voids. This profile follow a similar but steeper trend to that follow by halos
in voids. 

\end{abstract}

\begin{keywords}
{large-scale structure of Universe; cosmology: observations; methods: statistical }
\end{keywords}

\section{Introduction}

There are many tests to constrain the models of structure formation,
which range from the large-scale statistics such as correlation
functions \citep{Peebles1980,Davis1983,Zehavi2002,nor02} or the power
spectrum \citep{Peebles1980, Tegmark,cole05}, to detailed studies of the
physical properties of individual galaxy clusters and voids.  Although
dense environments (i.e., galaxy clusters and groups) have been 
extensively studied, the underdense regions like giant voids attract
less attention. Yet, they are not less important and can provide
important information on galaxy formation
\citep{Hoyle2005,Goldberg2005,Rojas2005,darren05} and can give independent
constraints on cosmological models
\citep{Peebles2001,Manolis,croton,colberg,Solevi,charlie05}. Large voids have woken up
more and more interest since their first detections 25 years ago
(Kirshner et al. 1981, Rood 1981).  At present, thanks to the advent
of larger redshift surveys like the 2dFRGS \citep{colless01} and the SDSS \citep{york}, higher
resolution of cosmological simulations and better analytical
frameworks, we can extract accurate statistical information about
voids.  This information can be used in different ways but one of the
most important is to test the models of structure formation.

Voids can be studied in different ways.  One of the classical methods
is the Void Probability Function \citep[VPF,][]{White79,Fry1986},
which gives the probability that a randomly located sphere of a given
radius contains no galaxies (see e.g. Einasto et al 1991, Croton et al
2004, Solevi et al 2005). The number density of voids with radius
greater than $R$ is another useful statistic. This number density and also
the void significance can be estimated analytically (see Patiri et
al. 2004). So far, it has been done only using numerical simulations
or mock catalogs \citep{colberg}.

Voids in the 2dFGRS were previously studied by \citet{hoyle} and \citet{croton}.
\citet{croton} have measured the VPF for volume-limited galaxy samples covering the
absolute magnitude range \smash{$M_{b_{\rm J}}-5\log_{10}h=-18$} to $-22$. Their work
mainly  focused on the study of the dependence of the VPF on the moments of galaxy
clustering as a test to discriminate among different clustering models. They found that the VPF
measured from the 2dFGRS is in excellent agreement with the paradigm of hierarchical
scaling of the galaxy clustering. In addition, they showed that the negative binomial model gives a
good approximation of the 2dF data over a wide range of scales. On the other hand,
\citet{hoyle} have also calculated the VPF in the 2dFGRS obtaining similar results. They
have obtained the VPF for the dark matter matter halos in $\Lambda$CDM simulations and galaxy
mock catalogs from semi-analytic models of galaxy formation to compare with the data. They have
found that the results from the semi-analytic models that include feedback effects provide a VPF
that  agree with the VPF measured for the 2dFGRS and differ from that measured from the dark matter
distribution.

In spite of the fact that the notion of voids is not new, there is no
standard definition of what is a void. ``Voids'' sometimes mean quite
different objects. It all depends on used data and goals of the
analysis. For example, to explain the patterns of the galaxy
distribution in the Universe \citet{rien} and \citet{sheth} define
voids as irregular low-density regions in the density field.
\citet{colberg} use a similar definition to study void properties in a
$\Lambda$CDM universe. However, as these definitions are not based on
point distributions, it may be difficult to deal with the galaxy samples
provided by large scale redshift surveys. \citet{ElAd1997}, \citet{HoyleVogeley} and
\citet{hoyle} define voids as irregular regions of low number-density of
galaxies, which may contain bright galaxies. Thus, by
construction, voids are not empty even of very luminous and likely
massive galaxies. By contrast, \citet{gottlober} 
define voids as spherical regions which do not have massive objects (halos in
this case). Voids also can be defined in a statistical point of view as maximal spheres
\citep{otto,Einasto1989,gottlober,patiri}. 

In this paper we define voids as the maximal non-overlapping spheres 
empty of objects with mass (or luminosity) above a given value
. For example, we could define voids as maximal spheres empty of Milky
Way-size galaxies. While voids are empty of these galaxies, they could
have fainter galaxies inside. See Figure 1 for a graphical
representation of our definition.

The first step to follow using voids as a test for large scale
structure and galaxy formation models is to develop a robust algorithm
to detect them, to calibrate it, and to obtain the statistical
properties of voids for different catalogs both real and simulated
ones. With this information and with the predictions made through the
analytical formalism we may be able to contrast different structure
formation models. To achieve this goal we  develop two
algorithms. They are conceptually different but are based on the same
definition of void. We develop them as complementary tools. One
algorithm is intended to search for all the voids in  galaxy or dark
matter halo samples and the other is developed to search for the
rarest voids.

Once we have the tools to detect the voids, we study statistical 
properties that will be used to test the structure formation 
models. These properties go from the void correlations to the 
redshift dependence of voids. In the present work, we apply 
the tools we mentioned above in order to study the statistic 
of voids in the 2dF Galaxy redshift Survey.
One important point that could provide 
clues on the galaxy formation processes is the galaxy contents of 
voids. In this work, we present the first results on the 
distribution of faint galaxies in nearby voids.

In section 2 we briefly describe how to detect voids. 
In section 3 we present the statistics of voids found in cosmological
numerical simulations and compare the performance of our algorithms.
In section 4 we present the voids that we have found in the 2dFRGS
together with their statistical properties. In section 5 we develop a
method to get from the redshift space coordinates, the real space
ones. In section 6 we present the statistics of the galaxies inside
nearby rare voids. finally, in section 7 conclusions and discussions
are presented. In Appendix A we give details of our void finders presented in section 2. 
Here we also provide tests in order to check the performance of them.

Throughout this work we adopt a \LCDM~ cosmology model with parameters 
$\Omega_m=0.3$ and $\Omega_{\Lambda}=0.7$.

\begin{figure}
\includegraphics[width=\columnwidth]{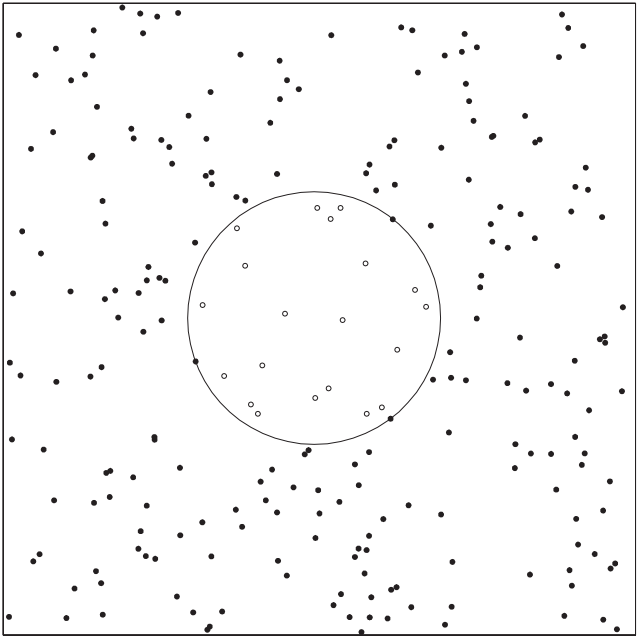}
\caption{A graphic representation of our void definition. Voids 
are maximal non-overlapping spheres, in the observational domain, which
are empty of objects classified by some intrinsic property. For example, in 
galaxy samples our voids will be defined by galaxies more luminous than a 
given luminosity $L$ (filled circles). Open circles denote galaxies 
with luminosity fainter than $L$. Note that for simplicity we do not show 
the open circles located outside the maximal sphere.}
\label{fig:fig1}
\end{figure}

\section{Void Detection Algorithms}
\label{sec:detect}

Once we have defined what is a void (as the maximal non-ovelapping spheres empty
of galaxies brighter than a given magnitude), the next step is to develop
an algorithm to be able to find them in galaxy or dark matter halo samples. 
The computation structure of the algorithm naturally will depend mainly on the 
void definition. Following the definition we have stated in the previous section, the 
algorithm should try to find the maximal sphere that the void can accommodate.

There are many algorithms in the literature inspired by the different
void definitions (see e.g. Einasto et al 1989, Kauffman \& Fairall
1991, El Ad \& Piran 1997, Aikio \& Mahoenen 1998, Gottl\"ober et
al. 2003, Colberg et al 2004). Aikio \& Mahoenen 1998 and Colberg el
al 2004 have developed similar algorithms. Following their definitions
of voids, they generate a field smoothing the point distribution on a
cubic grid and find the local minima; then, they put spheres around
those minima filling the entire underdense region. The radius of these
voids are the effective ones taken from the sphere that contain the
same volume as the void.  Gottl\"ober et al. 2003 have developed an
algorithm based on the minimal spanning tree for halos selected by
intrinsic properties, in this case with mass above a given value.
Although they have searched for voids as local maximal spheres, they
did not do voids statistics in their work.

Another algorithm in literature is the so-called {\em void finder}
developed by El Ad \& Piran (1997) and its modifications done by Hoyle
\& Vogeley (2002). This algorithm search for arbitrarily shaped
regions delimited by the so-called wall galaxies in order to get the
maximum volume of the void. They classify the galaxy sample in wall
galaxies and field or voids galaxies by mean of a criteria that
depends on the galaxy distribution itself, i.e. they define a length
parameter $l_n$ such that any galaxy that does not have $n$ neighbours
within a sphere of radius $l_n$ is classify as field galaxy. Note that
field galaxies could be, for example, bright galaxies. These field
galaxies are removed before searching for voids. Once the
classification is done, the algorithm search for the maximal spheres
defined by the wall galaxies based on a cubic mesh. Once all the
spheres are obtained, they define an overlapping parameter to
discriminate if two spheres belong to the same void. Similarly to
Aikio \& Mahoenen (1998) and Colberg el al (2004), they finally fill
the voids with spheres to get the maximum volume. Again, they define
the effective radius of a void by means of a sphere that contain the
same volume as the void.

Note that, although this may be interesting for some studies (e.g. the
shapes of voids), these kind of statistics degrade in some amount the
information available in the actual galaxy distribution, so they might
not be particularly powerful to conduct accurate statistical
inferences. This is similar to what happens with the binned data: the best 
statistical test using binned data is never better and 
usually worst than the best test using the raw data.

To identify voids, we designed two algorithms that are
conceptually different but both are based in our definition of
voids. The algorithms are complementary. They help us to investigate
the variety of aspects present in the statistics of voids. 

One of the algorithms, which we call {\em CELLS void finder}, was designed
to search for all the voids in a galaxy or halo sample based on a 
computational grid. This grid define the working resolution. 
To determine the void centers, the code computes the distances 
between each of the empty grid cells and all the galaxies or halos in the
whole observational domain, keeping the minimum distance. Once we have the list
with the minimum distances, we search for the local maxima which 
corresponds to the void centers. Obviously, the voids radius are those 
maximum distances.

The other algorithm, which we denominate {\em HB Void Finder} is conceptually simple. 
This code searches for the maximal non-overlapping spheres with radius {\em larger} than a given 
value. First of all, we generate over the sample a large sample of random spheres of a given radius. 
After this, we check and keep the spheres that are empty of galaxies. We inflate these spheres until 
they reach the maximum radius. Finally, we eliminate the overlapping spheres keeping the maximal ones. 
This code is very accurate and computationally efficient to search the biggest voids in a galaxy sample. 
Detailed descriptions of the two algorithms are given in the Appendix.

\section{Void statistics in simulations}
\label{sec:voids}

\subsection{Numerical simulations}

We perform a series of numerical simulations with the 
Adaptive Refinement Tree code (ART, Kravtsov et al. 1997) and the
Tree-SPH code GADGET (Springel, Yoshida \& White, 2001). Dark matter
halos are identified in the simulation by the Bound Density Maxima
algorithm (BDM, Klypin \& Holtzman 1997; Klypin et al. 1999).

In this work we detect and study voids in simulation boxes of $80
\mpch$, $120 \mpch$ and $500 \mpch$ size. The parameters of all the
simulations are summarized in Table 1. With these boxes we have enough
volume to study accurately the void statistics and compare the results
obtained with our two void finders.

\begin{table}
\caption{Parameters of simulations}
\begin{flushleft}
\label{tab:simu}
\begin{tabular}{cccc} \hline
Box  & Mass resolution & Number of particles & \sige \\
(\mpch) &   (\Msunh) &   &   \\ \hline
 80 & $ 3.18 \times 10^{8}$  & $512^{3}$  & 0.90  \\
120 & $ 1.07 \times 10^{9}$  & $512^{3}$  & 0.90  \\
500 & $ 7.80 \times 10^{10}$ & $512^{3}$  & 0.90  \\

\end{tabular} 
\end{flushleft}
\end{table}

\subsection{The statistics}

\begin{figure}
\includegraphics[width=\columnwidth]{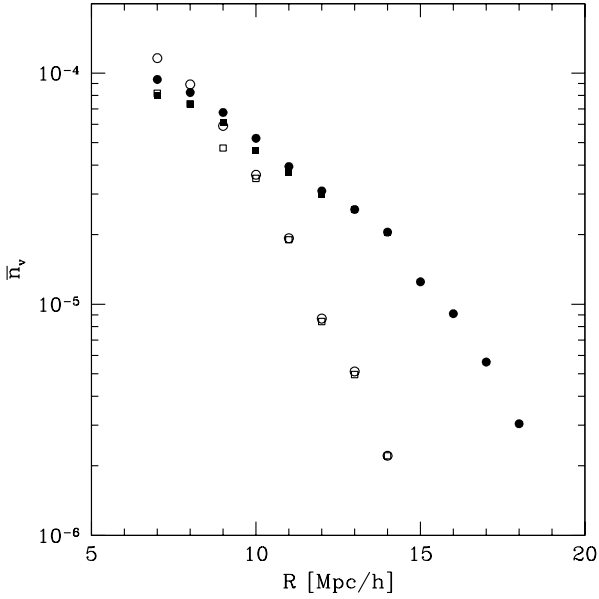}
\caption{Number density of voids in numerical simulations obtained with our two void finders (squares for the {\em CELLS void finder} 
and circles for the {\em HB void finder}). These results were obtained using our $80 \mpch$ and $120 \mpch$ boxes. 
The open circles and squares denotes the number density of voids defined by halos with masses larger than $1 \times 10^{12} \msun$ while
the filled circles and squares are the number density of voids defined by halos with masses larger than $5 \times 10^{12} \msun$. 
Here, we can see that both algorithms get the same results as voids are bigger (symbols are overlapped). However, as we go to more common voids the differences between 
codes are bigger. This is due to the fact that the {\em HB void finder} is not so efficient detecting common voids (see text for details).}
\label{fig:fig4b}
\end{figure}

In Figure \ref{fig:fig4b} we show the number density of voids as a function of their radius 
obtained with our void finders (squares for {\em CELLS void finder} and circles for the {\em HB void finder}). 
These results were obtained for voids defined by two halo masses ($1 \times 10^{12} \msun$ (open symbols) and 
$5 \times 10^{12} \msun$ (filled symbols)).

In order to search for voids with the {\em HB Void Finder}, we have 
generated $1 \times 10^{7}$, $2 \times 10^{7}$, $5 \times 10^{7}$ trial spheres
for the  $80 \mpch$, $120 \mpch$ and $500 \mpch$ boxes respectively.
In the case of the {\em Cell Void Finder} we defined a working resolution of $0.5 \mpch$ for 
the $80 \mpch$ and $120 \mpch$ boxes and $1.2 \mpch$ for the $500 \mpch$ box.

The obtained statistics shows that both algorithms gives very similar results. 
These results are indeed very useful to learn about the performance of
the void finders. Both algorithms give exactly the same results for the 
largest voids (these are rare voids), being the {\em HB Void Finder} 
the fastest and more precise. However, as we go to more common voids the
{\em Cell Void Finder} has the best performance. The small differences between
both codes for smaller voids are due to the fact that we need more realizations of the 
trial spheres in the {\em HB Void Finder}. As it is expected, the
voids are more numerous and larger the more massive the halos defining the void are.

%\begin{figure}
%\includegraphics[width=\columnwidth]{numsimf.eps}
%\caption{Voids in numerical simulations. We show a 2-D projection of
%the maximal spheres of radius greater than $9 \mpch$ in a $20 \mpch$ slice of 
%the $120 \mpch$ simulation box. Filled dots are dark matter halos with 
%virial mass greater than $2 \times 10^{12}\Msunh$. Note that those halos inside the maximal spheres 
%are due to projection effects.}
%\label{fig:fig4}
%\end{figure}

\section{Void statistics in the 2\lowercase{d}F Galaxy Redshift Survey}

\subsection{The galaxy samples}

In the present work we use the 2dFGRS final data release (Colless et al. 2003) 
to obtain the void statistics in large galaxy redshift samples. 
The source galaxy catalogue of the 2dFRGS is taken from the APM galaxy 
catalogue (Maddox et al. 1990). The spectroscopic targets are galaxies with 
extinction corrected magnitudes brighter than $b_{J}=19.45$. The median 
depth of the survey is $z \sim 0.11$. The final data releases contains a 
total of 221,414 high quality redshifts. There are two large contiguous survey 
regions, one in the south Galactic pole (SGP) and another one towards the north 
Galactic pole (NGP). There are also a number of random fields which we have 
eliminated from our void search. Full details of the 2dFGRS can be found in 
Colless et al.(2001, 2003).

In order to search for the voids we have selected from 2dFGRS 
two rectangular regions: the region in the SGP defined by 
$-34^{\circ} 40^{\prime} < \delta < -25^{\circ} 12^{\prime}$
and $21^{\rm h} 49^{\prime} < \alpha < 3^{\rm h} 26^{\prime}$ and 
the region in the NGP defined by 
$-4^{\circ} 35^{\prime} < \delta < 2^{\circ} 17^{\prime}$ 
and $9^{\rm h} 33^{\prime}  < \alpha < 14^{\rm h} 54^{\prime}$; that is,
$\sim 690$ and $\sim 550$ sq. degrees respectively.

The 2dFGRS is magnitude-limited, i.e. the survey has been constructed by 
taking spectra of galaxies brighter than a fixed apparent magnitude 
of $b_{\rm J}=19.45$. However the survey is homogeneously complete
up to 90\% at $b_{\rm J}$=19.0 (see Norberg et al. 2002). A magnitude-limited 
galaxy survey is not uniform in space, since intrinsically faint galaxies 
have been observed only if they are relatively nearby, while 
at large distances only bright galaxies will be targeted. 
This non uniformity of the magnitude-limited survey must 
be taken into account in order to make our void analysis. There are
mainly two ways to deal with this; one is to use
the selection functions provided by the 2dFGRS and another more
simple way which we have followed here is to build volume-limited samples.

We construct four volume-limited samples, two for each survey region 
(SGP and NGP), one with depth $D_{max} = 406.15 \mpch$ which corresponds to 
$z=0.14$. For this $D_{max}$ we have a limiting absolute magnitude
$M_{b_{\rm J}}^{lim}= -19.32 +5{\rm log} h$ and the other with depth $D_{max} = 571.71 \mpch$ 
corresponding to $z=0.2$ and a limiting absolute magnitude
$M_{b_{\rm J}}^{lim}= -20.181 +5{\rm log} h$. All distances are comoving ones. In Table 2 we give the properties of the 
volume-limited samples. Now, we have guaranteed that any galaxy brighter than 
$M_{b_{\rm J}}^{lim}$ is observed in our volume. We have computed the absolute 
magnitudes from the apparent magnitudes assuming a $\Lambda$CDM cosmology
and applying the needed corrections to model the change in the galaxy magnitudes
due to the redshifted $b_{\rm J}$ filter bandpass ({\em k}-correction) and to account for
the galaxy evolution ({\em e}-correction). These corrections for each galaxy are given in Norberg et al.(2002). 

\begin{table*}
\begin{center}
\caption{Parameters of our Volume Limited Samples.}
\label{tab:VLS}
\begin{tabular}{cccccc} \hline
Name   &     $M_{lim}$     & $z_{max}$  & $D_{max}$ &     Volume         & $N_{gxs}$ \\
       & $M_{b_{\rm J}} -5{\rm log}~ h$ &            &  $\mpch$  & $10^{6} h^{-3}\Mpc^{3}$ &     \\ \hline
 SGP1  &      -19.32       &   0.14     &  406.15   &     4.693          & 22037     \\
 SGP2  &      -20.181      &   0.20     &  571.71   &    13.088          & 14475     \\
 NGP1  &      -19.32       &   0.14     &  406.15   &     3.749          & 19695     \\
 NGP2  &      -20.181      &   0.20     &  571.71   &    10.456          & 11404     \\

\end{tabular} 
\end{center}
\end{table*}

To test the spatial homogeneity of our SGP1,2 and NGP1,2 volume-limited samples we have 
computed the average of the cube of the radial distances, i.e. a modified version of the $V_{max}$ test 
(Rowan-Robinson 1968), which for a 
homogeneous sample must satisfy:

\begin{eqnarray}
\big<\big(\frac{r}{r_{max}}\big)^{3}\big> = \frac{1}{2} \pm \frac{1}{\sqrt{12 N^{\prime}}} \\
N^{\prime}=\frac{N}{(1+N<\xi(r)>)} \nonumber
\end{eqnarray}

where $r$ is the comoving distance to each galaxy, $r_{max}$ is the same as 
$D_{max}$, the maximum distance of the sample, $N$ is the number of galaxies 
(given in table 2) and $<\xi(r)>$ is the average value of the correlation function 
over all pair of galaxy positions within the sample, defined as:

\begin{equation}
<\xi>=\frac{1}{V^2} \int\int \xi(|{\bf r_{1}}-{\bf r_{2}}|) ~d{\bf r_{1}}d{\bf r_{2}}
\end{equation}
where $V$ is the volume of the sample. Assuming for $\xi(r)$ (Peebles 1980):
\begin{equation}
\xi(r)=\big(\frac{r}{5.4 \mpch}\big)^{-1.77}  \label{eq:xi}
\end{equation}
we found $<\xi(r)>=6.7910 \times 10^{-3}$. It must be noted that we have used eq.(\ref{eq:xi}) for
any value of $r$, while in fact, it is known that for $r >> 10 \mpch$ $\xi(r)$ must be close to the
fourier transformed of the linear power spectra. This would lead to smaller values of $<\xi(r)>$. However,
this would not affect our analysis too much.
With this value we find that for a homogeneous sample:

\begin{equation}
\eta \equiv \big<\big(\frac{r}{r_{max}}\big)^{3}\big> = \frac{1}{2} \pm 0.024.
\end{equation}
The actual $\eta$ values for the SGP1 and NGP1 are $\eta(SGP1)=0.524$ and $\eta(NGP1)=0.51$. 
For the SGP2 and NGP2 to be homogeneous (assuming that $<\xi(r)>$ takes the same
values as in the previous sample):

\begin{equation}
\eta \equiv \big<\big(\frac{r}{r_{max}}\big)^{3}\big> = \frac{1}{2} \pm 0.0253,
\end{equation}
and the actual values are $\eta(SGP2)=0.511$ and $\eta(NGP2)=0.475$.
Therefore, we can conclude that we do not detect inhomogeneities.

\subsection{The void statistics}

\begin{figure}
\includegraphics[width=\columnwidth]{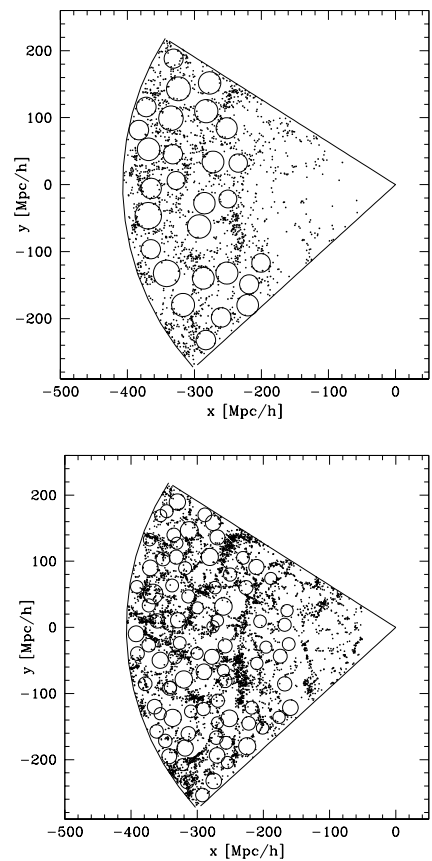}
\caption{Voids in the 2dFGRS. We show the maximal spheres  in the 
NGP1 volume-limited sample of galaxies (filled circles). The upper panel 
 shows the voids with radius larger than $7.5 \mpch$ defined by galaxies 
brighter than $M_{b_J}=-20.5 -5{\rm log}~h$. 
There are 2074 galaxies in this sample. 
In the lower panel we show the voids with radius larger than $13.0 \mpch$ defined by galaxies brighter than 
$M_{b_J}=-19.32 -5{\rm log}~h$. The number of galaxies is 19695. Some galaxies 
lie inside the maximal spheres due to projection effects.}
\label{fig:fig5}
\end{figure}

As described in section 3.2, once we have constructed our 
samples, we have to define which objects will define the voids.
In the SG1 and NGP1 volume limited samples we present in table 2 we 
search for voids defined by two types of galaxies. In the one hand, we search voids defined 
by galaxies brighter than $M_{b_{\rm J}}= -19.32 +5{\rm log} h$. Using these samples we  also find voids defined by %%@
galaxies brighter than $M_{b_{\rm J}}= -20.5 +5{\rm log} h$. The number of galaxies brighter than $M_{b_{\rm J}}= %%@
-20.5 +5{\rm log} h$ is 2427 for the SGP1 sample and 2074 for the NGP1. In the SGP2 and NGP2 samples, we have searched %%@
for voids defined by galaxies brighter than $M_{b_{\rm J}}= -20.181 +5{\rm log} h$. 
In Figure \ref{fig:fig5} we show a plot with the voids we have detected in the NGP1 sample. In table 3 we 
summarize the results for the NGP1 and SGP1 samples. Note that, as we saw in the simulations, voids
defined by bright galaxies are larger than the ones defined by fainter ones. In table 4 we present the 
statistics for the NGP2 and SGP2 samples.
We have obtained these voids applying the {\em HB algorithm}. In the radius ranges that we present here both codes have a %%@
similar performance. We generate $8 \times 10^{7}$ trial spheres to search for voids larger than $7.5 \mpch$ which are %%@
our main interest.

\begin{table}
\caption{Statistics of Voids in our NGP1 and SGP1 samples. $N_{NGP1}$ and $N_{SGP1}$ are the
number of voids larger than the given radius. We present the void statistics for two different
defining galaxies.}
\label{table4} 
\begin{tabular}{ccc}  

\hline  \hline
Radius & $N_{NGP1}$ & $N_{SGP1}$  \\ 
($\mpch$) &  &     \\
\hline  \hline

\noalign{\smallskip}
\noalign{$M_{b_{\rm J}}^{lim}=-19.32 -5{\rm log} h$}
\noalign{\smallskip}

\hline
 7.5 &136  & 220  \\
10.0 & 48  &  80  \\
12.0 & 11  &  25  \\		
13.0 &  6  &  11  \\
14.0 &  1  &   6  \\			 

\hline 

\noalign{\smallskip}
\noalign{$M_{b_{\rm J}}^{lim}=-20.5 -5{\rm log} h$}
\noalign{\smallskip}

\hline
13.0  & 28  & 43  \\
14.0  & 24  & 35  \\
16.0  & 14  & 26  \\	     
17.0  &  7  & 17  \\
19.0  &  2  &  6  \\

\hline 

\end{tabular}
\end{table}

\begin{table}
\caption{Statistics of Voids in our NGP2 and SGP2 samples.}
\label{table5} 
\begin{tabular}{ccc}  

\hline  \hline
Radius & $N_{NGP2}$ & $N_{SGP2}$  \\ 
($\mpch$) &  &     \\
\hline  \hline

\noalign{\smallskip}
\noalign{$M_{b_{\rm J}}^{lim}=-20.181 -5{\rm log} h$}
\noalign{\smallskip}

\hline
13.0 & 68  & 101  \\			 
15.0 & 36  &  53  \\
16.0 & 25  &  31  \\		
17.0 &  9  &  23  \\
18.0 &  5  &  14  \\
21.0 &     &   2  \\
     
\hline 

\end{tabular}
\end{table}

\begin{figure}
\includegraphics[width=\columnwidth]{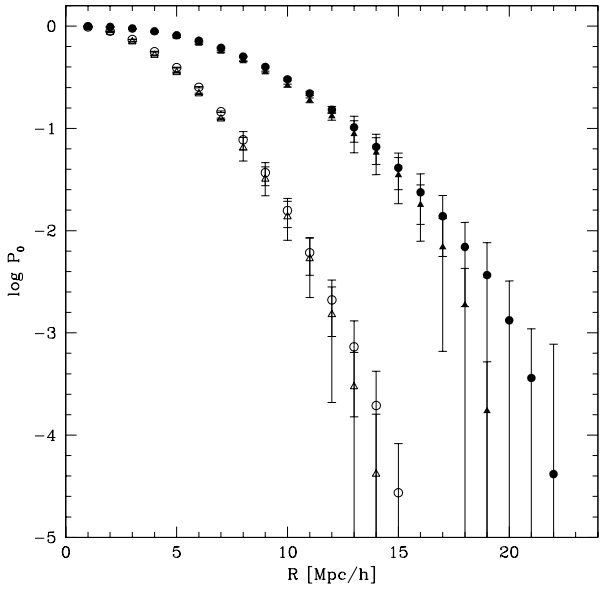}
\caption{Void Probability Function for the 2dFGRS. We show the VPF for the SGP1,2 samples
(open and filled circles respectively and also for the NGP1,2 samples (filled and open
triangles respectively).}
\label{fig:fig5vpf}
\end{figure}

We have also calculated the VPF which is the probability that a randomly
located sphere of fixed radius contains no galaxies. Figure \ref{fig:fig5vpf} 
presents our VPF, which we find for the SGP1,2 and NGP1,2 samples. 
We calculate the {\em rms} of the VPF in the following way.
Assuming that voids are independent, we get from the central limit theorem
that

\begin{equation}
rms^{2}(P_{0}(r))=\frac{N(r)}{V^{2}}(<\Delta v^{2}>-<\Delta v>^{2}) \label{eq:vpf1}
\end{equation}
where $\Delta v$ is the volume where the center of an empty sphere of
radius $r$ may be moved so that it remains empty. $V$ is the volume of the sample and
$N(r)$ is the number of voids found with radius larger than $r$ in that volume.
The parenthesis in the right hand side of eq.(\ref{eq:vpf1}) divided by $V^{2}$ is the contribution
of each individual void to the variance of the VPF.

On the other hand we have from \citet{Juan92} that
\begin{equation}
\frac{N(r)}{V}=\frac{P_{0}(r)}{<\Delta v>} . \label{eq:vpf2}
\end{equation}
Now we square this expression and substitute it in expression (\ref{eq:vpf1}) to obtain:

\begin{equation}
rms^{2}(P_{0}(r))=\frac{P_{0}^{2}(r)}{N(r)}~(g(r)-1) \label{eq:vpf3}
\end{equation}
\begin{equation}
g(r)\equiv \frac{<\Delta v^{2}>}{<\Delta v>^{2}} 
\end{equation}
So that, taking into account the correlations in the sample, the {\em rms} of the VPF is given by:
\begin{equation}
rms(P_{0}(r)) \simeq g(r)^{1/2}~\frac{P_{0}(r)}{N(r)^{1/2}}~(1-w\bar{n}_{v}(R))^{1/2}
\end{equation}
where $w\bar{n}_{v}(R)$ is as defined in Eq.(\ref{eq:eq4.4.1}).
In the rare voids limit, i.e. when $P_{0} <<< 1$ it is shown that \citep{Juan92}:
\begin{equation}
g(r) \simeq 9.20
\end{equation}

For voids with radius larger than $12 \mpch$ in the SGP1 and NGP1 samples and larger than 
$17 \mpch$ in the SGP2 and NGP2 samples, the asimptotic value $g(r)^{1/2}=2.86$ is a good 
approximation. For smaller values of $r$, $g(r)^{1/2}$ is somewhat smaller (no less than 
$\sim2$). So in figure (\ref{fig:fig5vpf}), where the asimptotic value of $g(r)$ have 
been used, the error may be slightly overstimated for small values of $r$. It must be noted, 
however, that the probability distribution of the fluctutations of the VPF around the mean 
is strongly non-Gaussian for small values of $N(r)$; most fluctuation being quite smaller 
than the $rms$ and a few of them being very large. So, to decide the compatibility between 
couples of measurements, this fact has to be taken into account. In the present case, however, 
this problem do not arize because all results are within the error bars)

Our results are in good agreement with the previous calculations of the VPF (Hoyle \& Vogeley 2004, Croton et al. 2004). 
In Croton et al.(2004) the VPF was obtained for the whole 2dFGRS while Hoyle \& Vogeley (2004) have studied the VPF
for both strips (NGP and SGP). In this last result, they found that the VPF for both strips are not
compatible for large spheres (see figure 7 in Hoyle \& Vogeley 2004). This is be due to the fact that 
they underestimated the error bars. We have shown above that using a more accurate expression for 
the {\em rms} of the VPF we obtain that both strips are compatible. 

\subsection{Voids spatial distribution}

If we neglect the redshift dependence of the power spectra, the spatial distribution
of the voids found in a statistically homogeneous sample must be statistically 
homogeneous itself, but conditioned to the fact that the maximal sphere must lay
within the sample. This fact implies that the center of the maximal spheres
can not occupy all the volume of the sample. The available volume, $V(R)$, for
the maximal spheres of radius $R$ is:

\begin{eqnarray}
V(R)&=&\int_{\frac{R}{{\rm sin}(\frac{\Delta\delta}{2})}}^{r_{max}-R}\int_{\delta_{0}}^{\delta_{0}+\Delta\delta}
\int_{\alpha_{0}+B(r,R,\delta)}^{\alpha_{0}+\Delta\alpha-B(r,R,\delta)} r^{2} {\rm cos}~\delta d\alpha~d\delta~dr \label{eq:eq4.3.1} \\
B(r,R,\delta)&=&{\rm sin}^{-1}\bigg(\frac{{\rm sin}^{-1}(\frac{R}{r})}{{\rm cos}\delta}\bigg) \nonumber
\end{eqnarray}
Eq.(\ref{eq:eq4.3.1}) may be considerably simplified by carrying out the integrals over $\alpha$ and $\delta$ keeping 
$B$ fixed (using its value at $\delta_{0}+\Delta\delta/2$). This is a very good approximation since $\Delta\delta << 1$. 
So, we could write:

\begin{equation}
V(R) \simeq \int_{\frac{R}{{\rm sin}(\frac{\Delta\delta}{2})}}^{r_{max}-R} P(r,R) r^{2} ~dr
\end{equation}
\begin{eqnarray}
P(r,R) &\equiv& \big[{\rm sin}\big(\delta_{0}+\Delta\delta-{\rm sin}^{-1}(\frac{R}{r})\big)-{\rm sin}(\delta_{0})\big] \times \nonumber
\\
       & \times & \big[\Delta\alpha-2~{\rm sin}^{-1}\big(\frac{{\rm sin}^{-1}(\frac{R}{r})}
{{\rm cos}(\delta+\frac{\Delta\delta}{2})}\big)\big] \label{eq:eq4.3.2}
\end{eqnarray}
$\Delta\delta$,$\Delta\alpha$ are, respectively, the widths of the strips in declination and 
right ascension, while $\delta_{0}$ is its southernmost declination. 
Although expression (\ref{eq:eq4.3.2}) for V(R) is not exact, its percentual error is completely negligible (less than $10^{-5}$).

Using $P(r)$ we may obtain the mean of the cube of the distance (from the observer), $r$, to the center of the maximal
sphere with radius larger than $R^{\prime}$:

\begin{eqnarray}
\eta(R^{\prime}) &\equiv& \big<\big(\frac{r}{r_{max}}\big)^{3}\big>_{R\geq R^{\prime}} = \\ &=&\frac{1}{V(\bar{R})}
\int_{\frac{\bar{R}}{{\rm sin}(\frac{\Delta\delta}{2})}}^{r_{max}-\bar{R}} P(r,\bar{R}) 
\big(\frac{r}{r_{max}}\big)^{3}r^{2} dr \nonumber
\end{eqnarray}
where $\bar{R}$ is the mean radius of all maximal spheres larger than $R^{\prime}$.

In a similar manner we may obtain the variance for the estimate of $\eta(R/\mpch)$ from the void list we
have obtained in the 2dFGRS. The values found in the SGP1,2 and NGP1,2 for voids defined by galaxies brighter than $-19.32 +5{\rm %%@
log}h$ are listed in table 5).

\begin{table}
\label{spa1} 
\begin{tabular}{ccc}  
\hline  \hline
Radius & $\eta$ & $\eta$  \\ 
($\mpch$) & NGP1 & SGP1    \\
\hline
 7.5 & $0.512~(0.563\pm 0.02)$  & $0.503~(0.53\pm 0.02)$ \\
10.0 & $0.579~(0.579\pm 0.038)$ & $0.508~(0.54\pm 0.03)$   \\
13.0 & $0.66(0.615\pm 0.116)$    & $0.433~(0.558\pm 0.08)$  \\	     

\hline 
     &  NGP2 & SGP2 \\

\hline
18.0  & $0.475~(0.611\pm 0.085)$ & $0.588~(0.556\pm 0.06)$  \\

\hline 

\end{tabular}
\caption{The values of $\eta(R^\prime)$ found in the SGP1,2 and NGP1,2 samples. 
The values within the parenthesis are $1\sigma$ predictions 
assuming that the center of the maximal spheres are uniformly
distributed conditioned to lay entirely within the sample (see text).}
\end{table}

From the analysis of these results we may conclude that the voids are essentially compatible with homogeneity.
There is, however, a slight, although statistically significant, trend to values lower than homogeneous that we will %%@
discuss in subsection (3.5).

Now, we will consider the nearest neighbour statistics which together with the void one point statistic we have %%@
studied above is sufficient to grant the validity of the statistical analysis in the following subsections.

The maximal spheres are chosen so that they do not overlap, in order that they very rarely corresponds to the same
connected underdensity. So, by construction, for the maximal spheres with radius larger than $R$ the two point %%@
correlation function of their centers is $-1$ at least to a distance of $2R$. The main question is how is the %%@
correlation at larger distances. Note that if more than one maximal sphere were associated with the same underlying %%@
connected underdensity the correlation would be positive for distances somewhat larger than $2R$. However, as there is %%@
typically only one maximal sphere per underdense region and in the standard scenario of structure formation these regions are %%@
essentially uncorrelated (for the relevant $R$ values)(Colberg et al. 2004), we do not expect to find correlations %%@
between the maximal spheres.

To test whether voids are correlated we compare the actual nearest neighbour statistics with the theoretical %%@
predictions corresponding to centers which are uncorrelated for distances larger than $2\bar{R}$ and completely %%@
anticorrelated for smaller distances. For maximal spheres with radius $R$ the theoretical predictions for the mean and %%@
quadratic mean distance to the nearest neighbour ($<D_{oo}>$ and $<D_{oo}^{2}>$
respectively) is given by:
\begin{equation}
<D_{oo}>=\bigg(\frac{4\pi\bar{n}_{v}(R)}{3}\bigg)^{-1/3}\int_{w_{0}}^{\infty} e^{-(w-w_{0})} w^{1/3} dw
\end{equation}

\begin{eqnarray}
<D_{oo}^{2}>&=&\bigg(\frac{4\pi\bar{n}_{v}(R)}{3}\bigg)^{-2/3}\int_{w_{0}}^{\infty} e^{-(w-w_{0})} w^{2/3} dw \label{eq:eqd15} \\
w_{0} &\equiv& \frac{4\pi}{3}(2\bar{R})^{3}\bar{n}_{v}(R) \nonumber
\end{eqnarray}
where $\bar{n}_{v}(R)$ is the mean number density of the maximal spheres larger than $R$ and $\bar{R}$ is the mean
radius. The sampling error of $<D_{oo}>$ when estimated using $N$ centers is:
\begin{equation}
rms(<D_{oo}>)=\frac{\sqrt{5}(<D_{oo}^{2}>-<D_{oo}>^{2})^{1/2})}{(3N-2)^{1/2}} \label{eq:eqd17}
\end{equation}
with these expressions we may compute the mean and the variance of $q$:
\begin{equation}
q\bigg(\frac{R}{\mpch}\bigg) \equiv (\bar{n}_{v}(R))^{1/3} E(<D_{oo}>)
\end{equation}
where $E(<D_{oo}>)$ is the estimate of $<D_{oo}>$ using the void sample. For typical voids we find for $q$:

\begin{eqnarray}
q_{SGP1}(7.5)&=&0.978 ~(0.958 \pm 0.006) \nonumber \\
q_{NGP1}(7.5)&=&1.029 ~(0.958 \pm 0.008) \nonumber \\
q_{SGP2}(13.0)&=&0.966 ~(0.879 \pm 0.01) \nonumber \\
q_{NGP2}(13.0)&=&1.1 ~(0.879 \pm 0.013) \nonumber
\end{eqnarray}
where $R=7.5 \mpch$ ($\bar{R}=9.58 \mpch$) correspond to SGP1, NGP1 and $R=13.0 \mpch$ ($\bar{R}=15.08 \mpch$) to
SGP2, NGP2. The values in the parenthesis correspond to $1\sigma$ predictions obtained using Eqns.(\ref{eq:eqd15}) and (\ref{eq:eqd17}). 

It is apparent that the estimated values are slightly shifted upwards. The bias is large in terms of the $rms$'s
but we can not conclude from this that the centers are anticorrelated since the predictions do not account for
border effect, which in this cases is small for the SGP samples but larger than the $rms$.

On the other hand, for rare voids, i.e. $R=13.0 \mpch$ for the SGP1, NGP1 samples ($\bar{R}_{SGP1}=14.14 %%@
\mpch$,$\bar{R}_{NGP1}=13.42 \mpch$) and $R=18.0 \mpch$ for the SGP2, NGP2 samples ($\bar{R}_{SGP2}=19.5 %%@
\mpch$,$\bar{R}_{NGP2}=18.45 \mpch$) we find:
\begin{eqnarray}
q_{SGP1}(13.0)&=&0.959 ~(0.716 \pm 0.004) \nonumber \\
q_{NGP1}(13.0)&=&1.398 ~(0.716 \pm 0.006) \nonumber   \\
q_{SGP2}(18.0)&=&0.971 ~(0.724 \pm 0.046) \nonumber   \\
q_{NGP2}(18.0)&=&1.071 ~(0.724 \pm 0.070) \nonumber
\end{eqnarray}
The upward bias with respect to the values for the uncorrelated centers (without accounting for border effects)
is now large even in the SGP. To asses the border effect we have simulated a random set of spheres with
$R=19.5 \mpch$ (the mean radius of the maximal spheres with radius $\geq 18.5 \mpch$ in the SGP2 sample)
with the condition that they lay entirely within the catalogue and do not overlap, then we obtain the nearest
neighbour statistics.
The average over several realizations of the mean nearest distance is:
\begin{equation}
\bar{n}_{v}(18.0)<\bar{D}_{oo}> = 0.88 \pm (0.05\pm \epsilon)
\end{equation}
which is quite larger than the value predicted without border effect (0.724). This is not enough, however,
to account for the measured value (0.971). The same result is found for $z_{max}=0.14$ since the border
effect in this case must be very close to that in the previous case. From these results we might conclude 
at least at the 99\% confidence level (being rather conservative about the value of $\epsilon$) that the center of the %%@
maximal spheres show some anticorrelation. We prefer,
however, to conclude simply that those centers are basically uncorrelated, which is the necessary result for
the subsequent statistical inferences, and leave open the question of the possible existence of a weak
anticorrelation, which shall be treated in more detail in a future work.
Incidentally, it must be noted that a slight anticorrelation of the centers arise 
naturally in the standard scenario, being compatible with the results of Colberg et al. 2004. The reason
is that the conexed underdense regions are somewhat larger than the maximal spheres they contain.

\subsection{Compatibility between SGP and NGP samples}

Here we compare the void statistics found in our samples in order to asses their compatibility.
To this end we must estimate the mean number density of voids, $\bar{n}_{v}(R)$, in each sample
and obtain their sampling error. To estimate $\bar{n}_{v}(R)$ we simply divide the number of voids,$N(R)$,
with radius larger than $R$, by the mean available volume $V(\bar{R})$. To estimate the sampling
error we must take into account that, although the center of the maximal spheres are basically uncorrelated
for $r > 2\bar{R}$, they are totally anticorrelated for $r < 2\bar{R}$ (we use a model in which all spheres
have radius $\bar{R}$). In this case the number of spheres in a given volume do not follow a Poissonian 
distribution, but a negative binomial one (Betancort-Rijo 1999). For this distribution the sampling error
of the estimate of $\bar{n}_{v}(R)$ could be obtained by means of the following expression:
\begin{equation}
rms(\bar{n}_{v}(R))=\frac{\bar{n}_{v}(R)}{(N(R))^{1/2}}(1-w\bar{n}_{v}(R))^{1/2} \label{eq:eq4.4.1}
\end{equation}
\begin{eqnarray}
w \equiv \frac{32\pi}{3}\bar{R}^{3} &\big[&1+2.73\bar{n}_{v}(R)\frac{32\pi}{3}\bar{R}^{3} \nonumber \\ &\times %%@
&(1-\frac{3}{4}
(\frac{\bar{n}_{v}(R)^{-1/3}}{2\bar{R}})+\frac{1}{8}(\frac{\bar{n}_{v}(R)^{-1/3}}{2\bar{R}})^{2})\big]^{-1} \nonumber
\end{eqnarray}
For SGP1 we find:
\begin{eqnarray} 
\bar{n}_{v}(7.5)&=&9.364 \pm 0.164 \times 10^{-5} \nonumber \\
\bar{n}_{v}(10.0)&=&3.98 \pm 0.11 \times 10^{-5} \nonumber \\
\bar{n}_{v}(13.0)&=&7.26 \pm 1.70 \times 10^{-6} \nonumber
\end{eqnarray}
For NGP1 we find:
\begin{eqnarray} 
\bar{n}_{v}(7.5)&=&9.597 \pm 0.214 \times 10^{-5} \nonumber \\
\bar{n}_{v}(10.0)&=&4.27 \pm 0.15 \times 10^{-5} \nonumber \\
\bar{n}_{v}(13.0)&=&6.56 \pm 2.10 \times 10^{-6} \nonumber
\end{eqnarray}
The corresponding $\bar{R}$'s are: $\bar{R}(7.5)=9.58$ and $\bar{R}(10.0)=11.35$ (the mean over
NGP and SGP), $\bar{R}(13.0)=14.14$ in the SGP1 and $\bar{R}(13.0)=13.41$ in the NGP1.

It is apparent that the three pair of values are compatible. It is true that for the SGP1 the
galaxy density is roughly a 10\% smaller than for the NGP1 and that the theoretical expression
\citep{patiri} predict about a 20\% enhancement of the density of voids with $R \geq 13 \mpch$ in the SGP1 over
that in the NGP1, and smaller differences for more common voids, but this differences do
not show up above the sampling errors.

Similar results were found for the SGP2:
\begin{eqnarray} 
\bar{n}_{v}(13.0)&=&1.685 \times 10^{-5} \nonumber \\
\bar{n}_{v}(15.0)&=&9.49 \times 10^{-6} \nonumber  \\
\bar{n}_{v}(18.0)&=&2.978 \pm 0.58 \times 10^{-6} \nonumber
\end{eqnarray}
and NGP2,
\begin{eqnarray} 
\bar{n}_{v}(13.0)&=&2 \times 10^{-5} \nonumber \\
\bar{n}_{v}(15.0)&=&1.202 \times 10^{-5} \nonumber \\
\bar{n}_{v}(18.0)&=&2.22 \pm 0.69 \times 10^{-6} \nonumber
\end{eqnarray}
  
Although the implications of this results for the large scale structure formation models
is left to for a future work, it is interesting to note that they are generally in good
agreement with the predictions of the standard model. 

\subsection{Redshift dependence of void number densities}  

We have checked that the galaxies in our samples are, within the sampling errors,
homogeneously distributed over the sample volumes. In this situation, 
the only possible explanation of an statistically significant departure of the 
void distribution from the homogeneity must be the redshift dependence 
of the properties of the large scale structure.

Using a theoretical framework \citep{patiri} to compute the number density of voids 
in the distribution of dark matter halos, we found 
that the predicted number density of voids larger than $13.0 \mpch$ for
$M_{b_{\rm J}}^{lim}=-19.32 +5{\rm log} h$ decreases by 28\% from z=0 to z=0.1 and a further 28\% when 
going to z=0.2. For $M_{b_{\rm J}}^{lim}=-20.181 +5{\rm log} h$ and $R \geq 18 \mpch$ the void density
decreases by 26\% from z=0 to z=0.1 and another 28\% from z=0.1 to z=0.2. 
Where, in order to link light with dark matter halos, we assumed that there is one galaxy in each dark halo. 
These numbers suggest the possibility of a measurable effect. 
To this end we conduct first a test where we divide every sample into two bins (a 'near' region
and a 'far' region) both with the same volume available for the voids considered.
Then we compare the total number of voids larger than the chosen radius in the
near regions with the total number in the far regions. If the
first number is significantly larger than the latter this should be interpreted
as evidence for a redshift dependence.

The problem now is that the effect we are trying to measure is strong only
for the rare voids which have poorer statistics. We reach a compromise
between both trends. We have chosen voids with radius ($R_{min}$) larger than $13 \mpch$ for the 
$z_{max}=0.14$ sample and $18 \mpch$ for the $z_{max}=0.2$ sample. 
In order to divide a sample into two subsamples with the same volume available for the centers of the
voids, we first obtain the mean radius ($\bar{R}$) of the voids larger than $R_{min}$, and, using Eq.(\ref{eq:eq4.3.1}) 
with $R=\bar{R}$, we obtain the availabe volume of the whole sample (which has already been done in
section 4.3). Then, we use again (\ref{eq:eq4.3.1}) with $R=\bar{R}$ but replacing $r_{max}$ with $r_{lim}+\bar{R}$ 
and search for the value of $r_{lim}$ giving for $V$ half the value corresponding to the whole sample. If voids were
uniformly distributed their centers would lie above and bellow $r_{lim}$ with equal probability. We have
explicitly checked this fact with numerical simulations.
For the sample SGP1, we have chosen $R_{min}=13 \mpch$, so $\bar{R}=14.14 \mpch$ which imply $r_{lim}=371.7\mpch$. 
9 voids were found with their centers bellow this distance and 2 above. Proceeding in a
similar way with the other samples, we found 3 voids closer than $349.1 \mpch$ 
and 3 voids further for the NGP1. For the SGP2 we found 5 voids closer 
than $475.0 \mpch$ and 8 further, while for the NGP2 we found 4 voids 
closer than $490.2 \mpch$ and 1 further. So, in total we have 21 voids in near regions 
and 14 in the far regions.

With these numbers we can infer at least at the 88\% confidence level
the presence of a redshift dependence. However, this is only a marginal
evidence. To make more patent this evidence we use a 
more efficient test. This test uses the $\eta$ values obtained in
4.3 which are shown in table 5. We use the $\eta$ values correpsonding to voids larger than $13 \mpch$ in
the NGP1 and SGP1 samples, and voids larger than $18 \mpch$ in the NGP2 and SGP2 samples. From each of
those four $\eta$ values we substract its expected value when uniformity is assumed (first number in the
parenthesis in table 5) and divide the result by the corresponding {\emph rms} (second number in the
parenthesis). Each of these quantities follow a Gaussian distribution with zero mean and variance 1
under the uniformity hipothesis. So, the sum of the four quantities must follow a Gaussian with zero
mean and variance 4 and the probability that these sum be smaller than the actually found result is:
\begin{equation}
1-\frac{1}{2}~erfc\bigg(\frac{\sum_{i=1}^{4}\frac{\eta_{i}-\bar{\eta_{i}}}{rms(\eta_{i})}}{2~\sqrt{2}}\bigg)=0.005
\end{equation}
So, we may infer at the 99.5\% confidence level the existence of non-uniformity on the void
distribution, that, given the uniformity of the galaxy distribution can only be explained by the
growth of density fluctuations as redshift decreases.
Note that the $rms(\eta)$ given in table 5 do not account for the anticorrelation between the
voids centers (when $r < 2R$). The last factor in expression (\ref{eq:eq4.4.1}) approximately accounts for this fact. 
If we were to use it, all the $rms(\eta)$ would be smaller and, consequently, we could reject the 
uniformity hypothesis at a larger confidence level. However, we prefer to give the conservative 
value quoted above since it has not any uncertaninty.

\section{Redshift Space Distortions}

Redshift galaxy maps like the 2dFGRS or SDSS are distorted by the peculiar 
velocities of galaxies along the line of sight. This effect, which produce deformations such as the ``finger of God'',
is due to the fact that the distance to the survey galaxies is obtained by means of the Hubble
distance which is obtained with the total velocity (the Hubble flow plus the peculiar velocity).

We present in this section a method to correct this effect for the galaxies around voids. We
first suppose that if in the distorted space we have a void of radius $R$, we will have
the same void of radius $R$ in the {\em real} space. This is a good approximation 
because the distortion around a void cause an elongation along the line of sight of the
maximal sphere without changing the ``transversal'' radius, i.e., although the distortion is not
volume conservative, the maximal radius will be approximately the same in both spaces.

With this assumption, we have in the position of the void in the distorted space a mean 
underdensity, $\bar{\delta}_{0}$, within the maximal sphere with radius $R$ \citep{patiri}:

\begin{equation}
\bar{\delta}_0=\int_{-1}^{\infty} \delta_{0} P(\delta_{0}/R) d\delta_{0}
\end{equation}
where $P(\delta_{0}/R)$ is the probability distribution for $\delta_{0}$ within a maximal sphere of radius $R$.

We associate with this void a matter distribution given by the mean profile, $\delta(r/\bar{\delta_{0}},R)$. 
In \citet{patiri} we derive this profiles.
So, it has a peculiar velocity profile $V(\delta(r),r)$ given by the spherical collapse model (Betancort-Rijo et al. 2005):

\begin{equation}
\frac{V(\delta(r),r)}{H}=\frac{-0.51}{3} \frac{r}{1+\delta} \delta_{l}(\delta) %%@
\big(\frac{d\delta_{l}(\delta)}{d\delta}\big)^{-1}
\end{equation}
where $H$ is the Hubble constant, $\delta_{l}(\delta)$ is given in Sheth \& Tormen (2002) and $\delta$ is the mean profile %%@
(note that for simplicity we do not show the dependence on $r$, $\delta_{0}$ and $R$). We have derived an
analytical approximation for $\delta(r/\bar{\delta_{0}},R)$ which gives good results for $r \leq 1.5R$:

\begin{equation}
\delta(r/\bar{\delta_{0}},R) \simeq \bar{\delta_{0}}+(0.1645+0.085\delta_{0})\big(\frac{r}{1.4R}\big)^{7}
\end{equation}
So that, we correct the galaxy ``measured'' distance $r(z)$:

\begin{equation}
r_{real} \simeq r(z)-
\frac{(\overrightarrow{x}-\overrightarrow{x}_{c}).\overrightarrow{x}}{|(\overrightarrow{x}-\overrightarrow{x}_{c})|
|\overrightarrow{x}|}\frac{V(\delta(r/\bar{\delta_{0}},R),r)}{H}
\end{equation}
where $\overrightarrow{x}$ is the vector to the galaxy and $\overrightarrow{x}_{c}$ is the vector to the center of the %%@
void. We apply this
correction to galaxies with distance to the center of the void $\leq 1.5R$. Once we have the corrected catalogue, 
we search for voids over this new catalogue.

In order to test our technique, we have applied it to a simulated catalogue, the Millennium Run galaxy catalogue 
(Springel et al 2005). The public available catalogue\footnote{It can be downloaded from 
http://www.mpa-garching.mpg.de/galform/agnpaper/} contains a total of about 9 million galaxies in the simulation box of $500 \mpch$. 
For each galaxy, it is available the position and velocity, the total and bulge galaxy magnitudes 
in 5 bands (ugriz SDSS bands), the total and bulge stellar mass, cold, hot and ejected gas mass, the
black hole mass and the star formation rate. The dark matter halos in the simulation were populated using semi-analytic models of 
galaxy formation (see Croton et al. 2005 for full details). We have constructed from the full simulated box a 
smaller one of $250 \mpch$. With this box we have enough volume to study large (and rare) voids. Also, as we have available in the original 
catalogue the coordinates in real space, we have constructed another box of $250 \mpch$ with the same galaxies 
but with the coordinates in redshift space.

We have applied our {\em HB void finder} to search for voids larger than $13.0 \mpch$ in both distorted
and real catalogues. With the list of voids in the distorted catalogue, we have applied the correction
to those galaxies that lie within $1.4$ radius of the voids. After we have obtained the corrected galaxy
catalogue, we run the void finder over this catalogue. In table (6) we show the statistics of 
voids in the 3 different catalogs: the distorted, the real and the corrected.

\begin{table}
\label{table:tb2} 
\begin{tabular}{cccc}  
\hline  \hline
Number &  &Radius [($\mpch$)]&  \\ 
of voids & Distorted & Real & Corrected    \\
\hline
 1 & 18.17  & 17.48  & 17.58 \\
 5 & 16.51  & 16.19  & 15.89 \\
10 & 15.70  & 15.09  & 14.93 \\      
15 & 15.42  & 14.57  & 14.58 \\
20 & 14.93  & 14.36  & 14.15 \\

\hline 

\end{tabular}
\caption{The void statistics found in the 3 catalogs generated from the millennium galaxy catalog: the 
redshift distorted, the real and the corrected. The first column show the number of voids found with 
radius larger than the values given in the 3 columns on the right corresponding to each catalog.}  
\end{table}

%In figure \ref{fig:fig6} we show a void of $16.37 \mpch$
%and its surrounding galaxies. In the first panel we show the distorted galaxies, in the second one
%the coordinates corrected galaxies and in third one the real galaxies. Note that the correction
%is very good.

From the study of the simulated catalogues we learn that, although the correction may be larger than
$3 \mpch$ (for $R \geq 16 \mpch$) for galaxies close to the line of sight to the center of the
voids, the correction for the radius of the maximal sphere is, as expected, much smaller. Even so,
the difference is not negligible for sufficiently rare voids. We find that the 20 largest voids in the
simulated catalogues are on average slightly smaller ($\sim 5$\%) in the corrected catalogue.

From the void statistics found in the corrected SGP1 and NGP1 samples we find that 
for the corrected SGP1 the ten largest voids are on average $0.83 \mpch$ smaller than
for the uncorrected one and that the number of voids larger than $12 \mpch$ in both samples is 25 for the uncorrected 
case and 14 for the corrected one. So, since the strongest constraint on the models comes from relatively
rare voids, it seems likely that the corrected catalogues must be used in order to be able to make
accurate inferences.

%\begin{figure*}
%\includegraphics[width=165mm]{dc.eps}
%\caption{Example of correction for redshift distortion. We show here
%an example of our technique to correct the redshift distortion applied
%to the mocks catalogues. In the left panel, we show a void of radius
%$16.37 \mpch$ in a slice of $8 \mpch$ and the surrounding galaxies 
%(with distorted coordinates) up to $4.0 \mpch$ from the surface of the sphere. 
%The central panel shows the same void but with corrected galaxy coordinates. In the right
%panel the real-space galaxy distribution is shown. The direction of the observer is
%parallel to the abscissas axis. The axes are in $\mpch$ units. Note the excellent
%agreement between the real and the corrected ones.}  
%\label{fig:fig6}
%\end{figure*}

\section{Galaxies in Nearby Voids}

\begin{figure*}
\includegraphics[width=165mm]{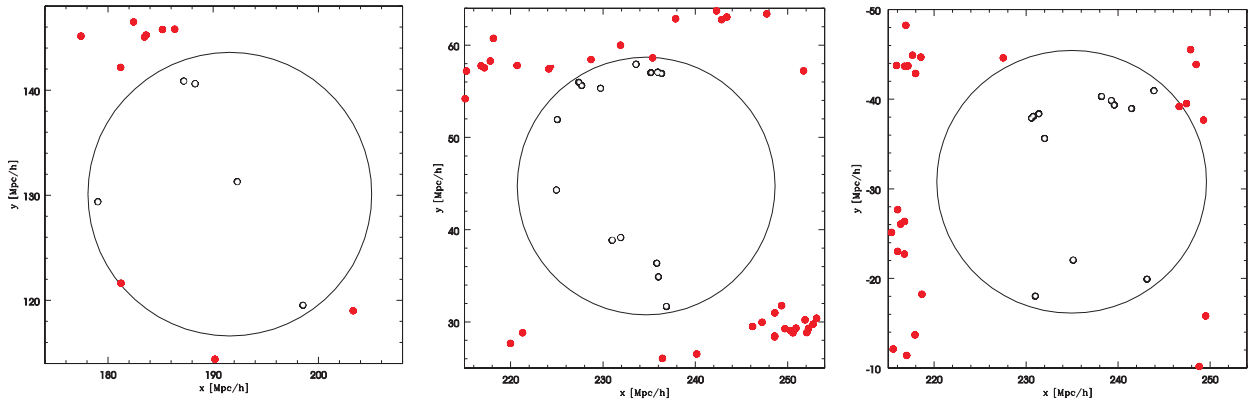}
\caption{Three examples of faint galaxies in voids. We show 
for each void a $6 \mpch$ central slab. In the left panel is shown a $13.5 \mpch$ 
void, in the central pannel a $14.0 \mpch$ void and in the right pannel a $14.7 \mpch$
void. The filled circles are the galaxies brighter than $-19.32 +5~{\rm log}h$ that 
define the voids. The open circles are the fainter galaxies inside the voids down to 
$M^{\rm lim}_{b_{\rm J}}=-18.3 +5~{\rm log}h$. Note that even though there is not a large
population of faint galaxies in these voids, the galaxies are not randomly distributed,
i.e. they are distributed in filamentry structures similar to those found in another scales
(see central and right panel).}
\label{fig:fig7}
\end{figure*}

\begin{figure}
\includegraphics[width=\columnwidth]{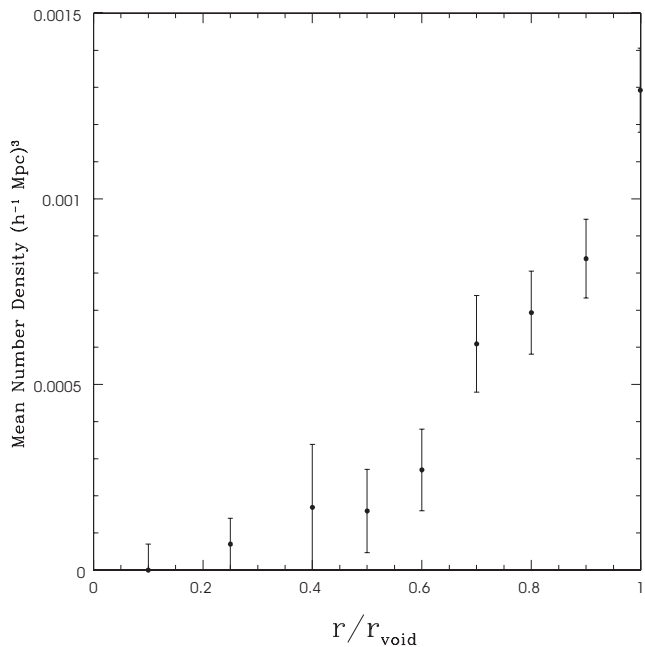}
\caption{Averaged enclosed number density profile of faint galaxies
($-19.32 +5~{\rm log}h < M_{b_{\rm J}} < -18.3 +5~{\rm log}h$) in
nearby voids as a function of the distance from the center of the void
(in void radius units). The mean number density of galaxies in the
field for these magnitude band is $7.0 \times 10^{-3} (\mpch)^{-3}$.}
\label{fig:fig8}
\end{figure}

In this section we study the galaxy content of the nearby rarest voids
in our 2dFGRS samples. To this end, we have selected the voids from
our SGP1 volume-limited sample described in Section 5.1 (see Table
3). The voids are defined by galaxies brighter than $-19.32 +5~{\rm
log}h$.  We have searched for faint galaxies down to $M^{\rm
lim}_{b_{\rm J}}=-18.3 +5~{\rm log}h$ inside nine voids with radius
larger than $13 \mpch$ in a volume limited sample up to z=0.095. These 
voids are uncommon (with mean radius of $14.0 \mpch$) due to the fact that a randomly placed
sphere with radius equal to the mean size of these voids should contain
about $50$ galaxies brighter than $-19.32 +5~{\rm log}h$.
In total we find 130 faint galaxies inside these 9 rare voids, i.e. on average 
14 galaxies per void. We have estimated the number density contrast of the 
galaxies located inside these voids by

\begin{equation}
\delta_{gal}=\frac{\bar{n}_{void}-\bar{n}}{\bar{n}},
\end{equation}
\noindent where $\bar{n}_{void}$ is the number density of galaxies inside the void 
and $\bar{n}$ is the number density of galaxies in the field. For the 
galaxies inside the voids we obtain $\delta_{gal}=-0.87$. In figure \ref{fig:fig7} 
we display a central $6 \mpch$ thick slice for three of these nine voids. We 
can see that, despite the fact that these voids are highly empty, the faint
galaxies populating them show interesting structures like filaments. 
Nevertheless, most galaxies are placed close to the borders of the voids
being their centers much emptier. Note that these galaxy patterns are similar 
to those found in voids in high resolution numerical simulations by Gott\"ober et al.
2003. In figure \ref{fig:fig8} we show the mean number density of faint galaxies in our rare 
voids as a function of distance to the void center (normalized to the void radius). 

\begin{figure}
\includegraphics[width=\columnwidth]{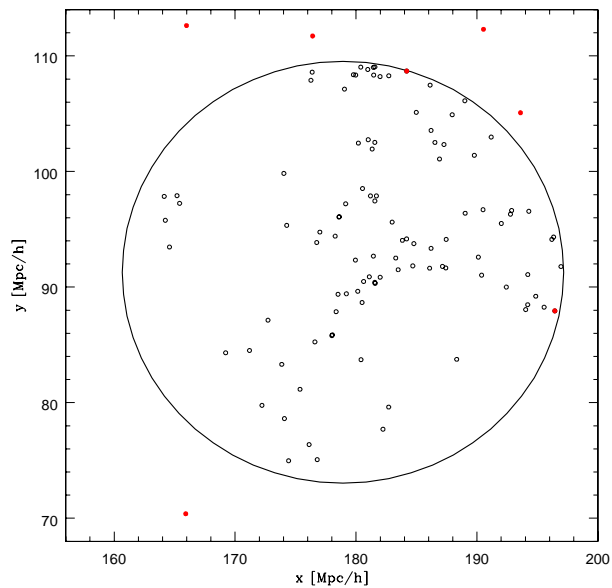}
\caption{Galaxies inside a common void. We show a central $8 \mpch$ thick slice of a $18.25 \mpch$ void defined by
galaxies brighter than $-20.5 +5~{\rm log}h$ (filled circles). The galaxies inside the 
void are fainter down to $M^{\rm lim}_{b_{\rm J}}=-18.3 +5~{\rm log}h$ (open circles).}
\label{fig:fig9}
\end{figure}

We have also studied the galaxies inside common voids, i.e. voids whose underdensities
are not too big. We have selected from the SGP1 sample the voids defined by galaxies
brighter than $-20.5 +5~{\rm log}h$. We found five voids larger than $15.0 \mpch$ up to 
z=0.095. For these radius, a randomly chosen sphere should contain only 3 galaxies 
brighter than $-20.5 +5~{\rm log}h$. In total there are 666 galaxies fainter 
than $-20.5 +5~{\rm log}h$ down to $M^{\rm lim}_{b_{\rm J}}=-18.3 +5~{\rm log}h$. 
These voids contain on average 10 times more galaxies than the rare voids discussed 
above, being the number density contrast of galaxies $\delta_{gal}=-0.54$. 
In figure \ref{fig:fig9} we show an example of a $8 \mpch$ thick slice of a $18.25 \mpch$ void in this sample. 
The galaxies inside this void almost fully fill the void (open circles).
This is not surprising due to the fact that the density contrast of these
voids is not too big.

\section{Discussion \& Conclusions}

We have developed two robust and accurate algorithms to detect non-overlapping maximal
spheres in halo or galaxy samples. We have applied them to several numerical simulations
boxes in order to study the performance of the algorithms. The results found were very
satisfactory.

We have then studied the void statistics in the 2dFGRS.
We have detected $\sim 350$ voids 
with radius larger than $7.5 \mpch$ defined by galaxies with $M_{b_{\rm J}} < -19.32 +5 {\rm log}h$
and $\sim 70$ voids with radius larger than $13.0 \mpch$ defined by galaxies with 
$M_{b_{\rm J}} < -20.5 +5 {\rm log}h$ in the volume limited samples up to $z_{max}=0.14$. We have
also obtained the void statistics for the volume limited samples up to $z_{max}=0.2$. For this case,
we have detected $\sim 170$ voids with radius larger than $13 \mpch$ defined by galaxies brighter
than $M_{b_{\rm J}} < -20.181 +5 {\rm log}h$.     

The number density of voids larger than $R$ found in the SGP and NGP samples are in good agreement
with each other for all values of $R$. We have obtained the VPF for both strips finding results
in good agreement with previous ones \citep{croton,hoyle}. We have shown, using an appropriate expression for the
VPF sampling errors, that the results found in both strips are statistically compatible. 

From the results of several statistical tests we conclude that except for the anticorrelation
implicit in the fact that the maximal sphere are chosen so that they do not overlap, they are
essentially uncorrelated. There is, however, some evidence for a weak additional anticorrelation,
which may be easely explained within the standard scenario of structure formation. We conclude
at least at 99.5\% confidence level that there is a dependence of the void number density on
redshift. We do this by means of a modified version of the $V_{max}$ test which reveals a small trend towards small z %%@
values.

We have also obtain preliminary results on the galaxy contents of nearby voids found in volume-limited galaxy
samples in the 2dFGRS. For the nine nearby voids up to $z < 0.095$ in our sample, we have found
inside them on average only 15 galaxies fainter than $-19.32 +5~{\rm log}h$ (the magnitude of the
galaxies which define the voids). These voids are rather empty compare to those defined by
brighter galaxies. The galaxies within the voids are not randomly distributed:
they are clustered forming well define filamentary structures as that observed in the large-scale 
structure of the galaxy distribution of the Universe. Moreover, this is the same pattern that show 
the dark matter halo distribution found inside voids in high resolution N-body simulations. 
The halos inside voids are distributed in a way that resemble a miniature of the Universe 
(see Figure 2 in \citet{gottlober}). We have also obtained the number density profile for
the galaxies inside the voids. The number density of faint galaxies fall almost a factor 7 from
the border of the voids to the inner half. Moreover, the number density of galaxies close of
the borders of the voids are still too low compared with the field (almost a factor 5).
\citet{gottlober} and \citet{patiri} have obtained the number density profile for halos in voids.
From the comparison of these results we can see that even though the number density of galaxies and halos follow
similar trends, the galaxy profile is steeper than halos (they fall just a factor 2). 

In a future work, we will use the framework developed in \citet{patiri} along with the
results given here for the number density of voids and their redshift dependence in order to constrain
the value of $\sigma_{8}$. Furthermore, using a halo occupation model (e.g. Berlind et al. 2003) along the lines
described in \citet{patiri} we shall establish constraints in the relationship between halos and
galaxies from void statistics and the statistics of the galaxies inside the voids. In particular,
we hope to be able to determine whether or not the conditional luminosity function depends on 
environment.

In the other hand, the study of the physical properties of the galaxies inside voids could imply
constraints on the galaxy formation processes. So, we will analyze the physical properties 
like colors, metalicities, star formation rates, etc. of galaxies in rare voids that are available from the 
biggest galaxy surveys (2dFGRS and SDSS) in order to test, for example, the galaxy luminosity function in voids and
the density-morphology correlations.

\section*{Acknowlegments}

We thank the referee for the comments and suggestions which improved the previous version of the paper. 
S.G.P. would like to thank the hospitality of the Department of Astronomy of the New
Mexico State University where part of this work was done.
We also thank support from grant PNAYA 2005-07789.

This work was also partially supported by the {\em Acciones Integradas Hispano-Alemanas}.
A.K. acknowledges support of NSF and NASA grants to NMSU.
Computer simulations have been done at CESGA (Spain), NIC J\"ulich, LRZ Munich and NASA.

The Millennium Run simulation used in this paper was carried out by the 
Virgo Supercomputing Consortium at the Computing Centre of the Max-Planck 
Society in Garching.

%\label{lastpage}

\appendix
\section[]{Void finder algorithms}
\subsection{Algorithm 1: {\em Cell} Void Finder}

The first step is to select the sample of galaxies or dark matter halos
in our redshift survey or cosmological simulation that will define the
voids. We will select galaxies with luminosity greater than a given value
$L$ or halos with virial mass greater than $M$.

Once the galaxy or halo sample has been selected our code generates a cubic mesh
where the cell size determines the working resolution. Afterwards, the objects
are assigned to cells. So we have three types of cells: the cells that 
contain galaxies or halos (filled cells), the cells that are 
empty but are inside the observational domain (observed empty cells) and those
cells which are also empty but they are located outside the observational 
domain or in a not observed region inside the observational domain (a 'hole')
(not observed empty cells). 
The code searches for the voids among the observed empty cells inside the 
observational domain. In the case of cosmological simulations the dark 
matter halo samples are in 3D boxes and generally all the cells are inside 
the observational domain. However, in the case of galaxy samples in redshift 
surveys which have irregular geometry, some of the cells are located 
outside the observational domain. We can easily determine  which cell has 
been observed and which has not by using the survey masks.

Once the cell classification procedure have been completed the code is 
then ready to search for the voids, finding their centers 
and radii. In principle, each observed empty cell could be a potential void center, 
but it is easy to realize that the observed empty cells which are located close to 
filled cells that contain galaxies or halos will not be the center of a void. 
So, in order to save computational time, 
we mark these neighbours cells and they will not be taken into account
at the time of searching for the void centers. This is an iterative process, 
i.e. once we have marked the observed empty cells closer to the filled cells we can 
go to the next level and mark the observed empty cells that are neighbours of 
already  marked empty cells. Note that we will stop this iteration depending 
on the working resolution (see Figure A1). 

To determine the void centers, the code computes the distances 
between each of the unmarked observed empty cell and all the galaxies or halos in the
whole observational domain and we retain the minimum distance. Once we have the list
with the minimum distances, we search for the local maxima which corresponds to the void centers. 
Obviously, the voids radius are those maximum distances.

Finally, the code removes the overlapping maximal spheres, keeping 
the biggest one, i.e. if the distance between two maximal spheres is
less than the sum of their radius, then the voids overlap and we remove
the smaller one.

The main advantage of the {\em Cell Void Finder} algorithm is that in only 
one run we get all the voids in the sample. However, its main disadvantage 
is that it consumes quite a lot of memory, which scale with the resolution 
that we require. There is a similar memory problem when
we have a large number of galaxies or halos in the sample.
If we are only interested in the biggest voids (rare voids) this 
algorithm is not the best strategy. Some studies
are focused in these kind of voids, so, here we have developed a 
complementary algorithm which is more efficient in this respect.

\begin{figure}
\includegraphics[width=\columnwidth]{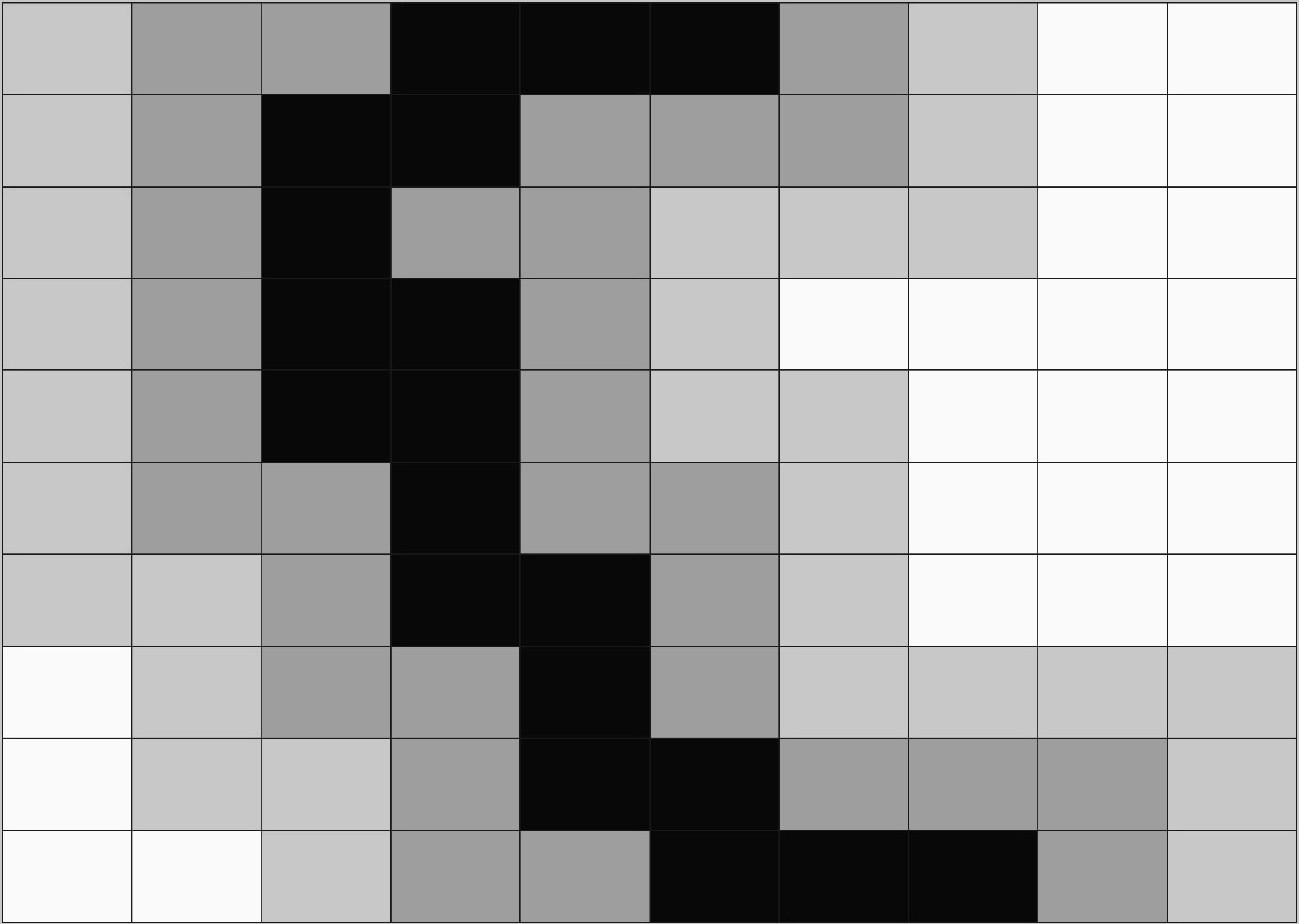}
\caption{An example of labeling of cells  for our {\em Cell Void Finder}. In order to 
save computational time, we mark the empty cells (dark grey cells) that are 
neighbours of those cells that contain objects (black cells). Note that this 
is an iterative process, i.e. we can now mark the empty cells that are 
neighbours of already marked empty cells (light grey cells). The stop of this 
iteration depends on the adopted resolution. The center of the void 
will be searched over the empty unmarked cells (white cells).}
\label{fig:fig2}
\end{figure}

\subsection{Algorithm 2: {\em HB} Void Finder}

Here we give the details of our second algorithm whose main task
is the detection of rare voids and we will use it as a complement of the CELLS algorithm.
The {\em HB Void Finder} is conceptually simple. It searches for 
the non-overlapping maximal spheres with radius {\em larger} than a given 
value. As we  mentioned above this code is designed 
for statistical studies focused on the biggest voids. In these cases the 
code is very accurate and computationally efficient as
compared with our {\em Cell Void Finder}.

The first step in the algorithm is to generate a sample of
random trial spheres with a fixed radius $R$. These spheres are
generated directly over the observational domain with
the condition that the entire sphere lies inside the 
observational domain. We check which spheres contains no objects
and keep them.

In the next step we find for each trial empty sphere the four nearest
object and we expand the sphere to contain them in its surface. These
new spheres are potentially maximal (see subsection A4).  Note that the new
expanded spheres could in principle contain objects. If this is the
case, those spheres are removed. The potential maximal spheres will be
{\em actual} maximal spheres if they do not overlap, which is decided
by means of the same criteria as for the {\em Cell Void Finder}
(i.e. if the distance between two spheres is less than the sum of
their radius they overlap, so we keep the largest one) and if the four
nearest objects are not located in the same hemisphere (see Figure A2).
We put in both algorithms the additional constrain that the maximal
spheres have to be entirely inside the observational domain.  Note
that, we need to do many realizations of the trial spheres in order to
get the maximal spheres. Typically, four realizations are needed for
each void in order to reach its maximum radius, when the maximal
sphere is only slightly larger than $R$ and an increasing number as
the maximal sphere is larger with respect to $R$.

\begin{figure*}[t]
%\centering
\includegraphics[width=165mm]{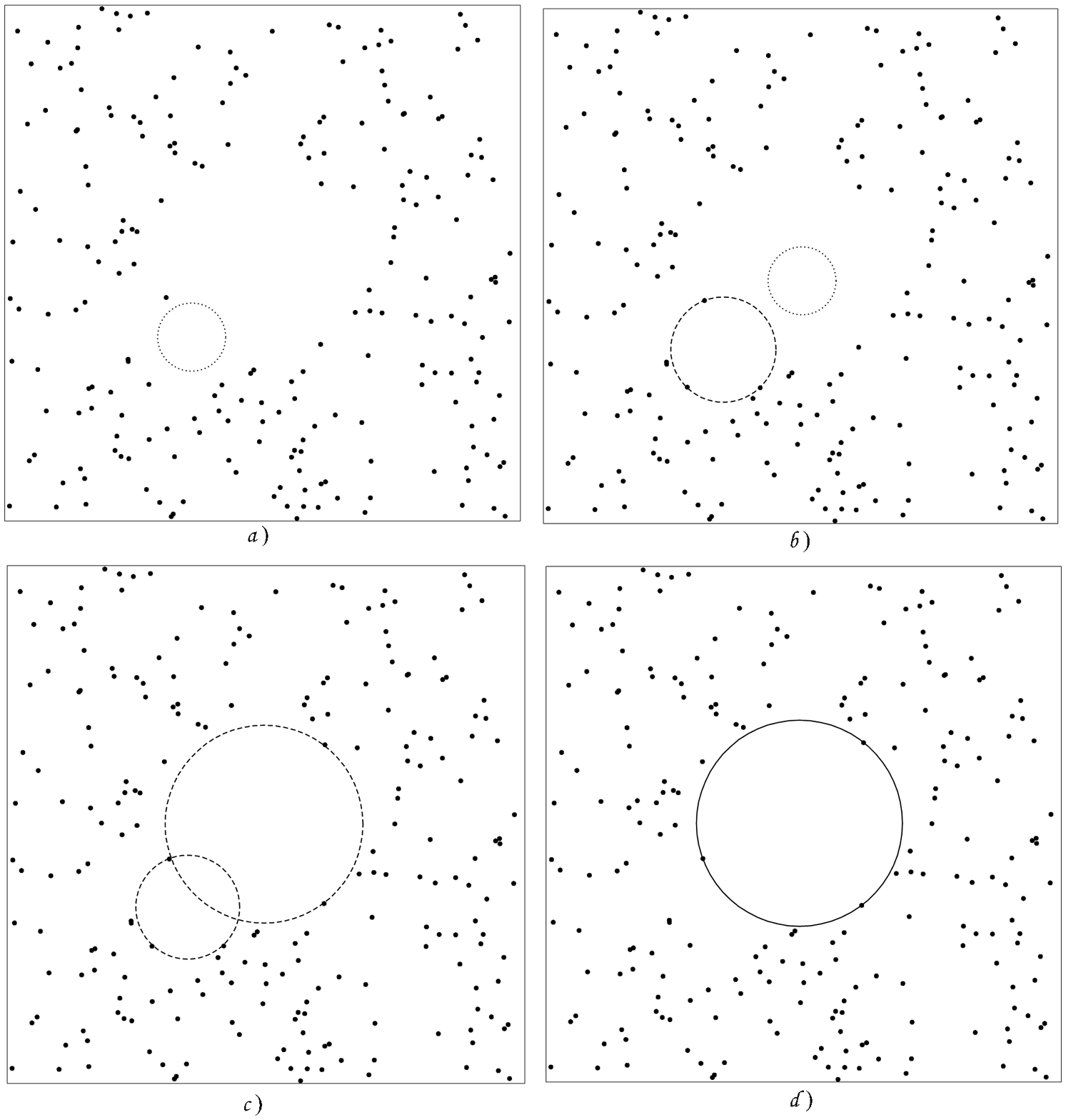}
\caption{Sequence of steps  in our \em{HB Void Finder}: $a)$ we throw a 
sphere of fixed radius (dotted circle) over the observed sample of objects,
$b)$ if this sphere is empty of objects defining the void (filled dots), 
we then grow the sphere (dashed circle) until its surface contains the 
four nearest objects (three in 2-D, see Appendix A). This is a potential 
maximal sphere. We throw another random sphere (dotted circle). Again, if this
one is empty, we obtain the potential maximal sphere (dashed circle in panel $c)$. 
$d)$ Finally, we check if the new maximal sphere overlaps with another one; 
if so, we keep the largest sphere. This is our void (filled circle). See text 
for details.}
\label{fig:fig3}
\end{figure*}

\subsection{Performance test}

We construct random samples of objects in order to test
the code performance and check the results of both algorithms. We 
generate two samples, one with 1,000 points and another one with 10,000 
points. Both samples are in a box of $100 \, \Mpc$. We give in Table A1
the void statistics computed from both algorithms. Note that the agreement 
between codes is excellent. 
We obtain the void statistics using $\sim 10^{7}$ trial 
spheres with the {\em HB Void Finder}. The resolution of the {\em Cell Void Finder}
is $0.5 \, \Mpc$.

The cpu time of the codes mainly depends on the number of particles, 
the number and radius of trial spheres in the case of the 
{\em HB algorithm} and on the number of cells (i.e. resolution) and 
the levels of neighbours cell marking for the {\em Cell Void Finder}. For 
example, in the case of the sample with 1,000 random particles and for the 
voids with radius larger than $10 \, \Mpc$, the {\em HB Void Finder} takes 
$\sim 1$ hour for $10^{7}$ trials, 
while the {\em Cell Void Finder} takes $\sim 3$ hours with a resolution of 
$0.5 \, \Mpc$. Notice that the voids with radius larger than $10 \, \Mpc$ 
are common in this box, so the running time differences are not so big 
between both algorithms. However, if we search for 
voids with radius larger than $12 \, \Mpc$, the {\em HB Void Finder} takes 
only 20 minutes, while the {\em Cell Void Finder} last
the same $\sim 3$ hours. These tests were done in a Pentium IV processor (3.06 GHz clock
and 2GB RAM) and in a Itanium-2 processor (1.5 GHz clock, 2GB RAM) giving both 
similar performances.

\begin{table}
\caption{Voids in random samples. We list the statistics of the maximal spheres
found by our algorithms in the sample of 1,000 and 1,0000 random points
in a Box of 100 $\Mpc$. $N_{HB}$ is the number of voids found with the 
\em{HB algorithm} with radius larger than the value given in the first column.
$N_{CELLS}$ is the same but for the voids found with the \em{Cells Void Finder}. 
See text for details.}
\begin{center}
\label{tab:voidrand}
\begin{tabular}{ccc} \hline
Sample  & Box:&100 Mpc   \\ 
 1000   &    &        \\ \hline
Radius  & $N_{HB}$ & $N_{CELLS}$  \\
$\Mpc$  &    &        \\ \hline 
  10.0  & 23 & 25     \\
  11.0  & 15 & 15     \\
  12.0  & 7  &  7     \\ \hline
10000   &    &        \\ \hline
  5.5   & 69 & 73     \\ 
  6.0   & 20 & 21     \\ 
  6.5   &  3 &  3     \\ 
  7.0   &  1 &  1     \\ 		

\end{tabular} 
\end{center}
\end{table}

\subsection[]{How to grow the trial spheres}

To determine the sphere passing through the four nearest objects to an empty trial sphere
we proceed as follow:
We first take the two nearest objects (whose coordinates we denote as 
$\overrightarrow{x}_{1}$ and $\overrightarrow{x}_{2}$) and calculate
the middle point $(\overrightarrow{x}_{1}+\overrightarrow{x}_{2})/2$. 
From this point, we move along a vector in the plane containing the three nearest objects and perpendicular to 
$\overrightarrow{x}_{2}-\overrightarrow{x}_{1}$ until we reach the point, 
$q$, where the distances to the third nearest object ($\overrightarrow{x}_{3}$) 
is the same as that to object $1$, then 
the distance between object 2 and $\overrightarrow{q}$ is also the same that 
the previous two (See figure A3).
Then we need to solve:
\begin{equation}
|\overrightarrow{x}_{1}-\overrightarrow{q}(w_0)| =|\overrightarrow{x}_{3}-\overrightarrow{q}(w_0)|
\end{equation}
where 
\begin{equation}
\overrightarrow{q}(w)=\frac{\overrightarrow{x}_{1}+\overrightarrow{x}_{2}}{2} + w\overrightarrow{j}
\end{equation}
and
\begin{equation}
\overrightarrow{j}=\frac{\overrightarrow{e}_{13}-(\overrightarrow{e}_{13}.\overrightarrow{e}_{12})\overrightarrow{e}_{%%@
12}}
{|\overrightarrow{e}_{13}-(\overrightarrow{e}_{13}.\overrightarrow{e}_{12})\overrightarrow{e}_{12}|}
\end{equation}
where
\begin{equation}
\overrightarrow{e}_{12} \equiv %%@
\frac{\overrightarrow{x}_{1}-\overrightarrow{x}_{2}}{|\overrightarrow{x}_{1}-\overrightarrow{x}_{2}|}; 
\overrightarrow{e}_{13} \equiv %%@
\frac{\overrightarrow{x}_{1}-\overrightarrow{x}_{3}}{|\overrightarrow{x}_{1}-\overrightarrow{x}_{3}|}
\end{equation}
with $w_{0}\in(-2R_{0},2R_{0})$, $R_{0}$ is the radius of the trial sphere.

Now, we repeat the same procedure described above but taking into account the fourth object, i.e. 
we move from $\overrightarrow{q}(w_0)$ perpendicularly to the plane of the figure A1 
until we reach the point $\overrightarrow{P}(t)$ where the distance between 
$\overrightarrow{x}_{4}$ and $\overrightarrow{P}(t)$ is the same as the distance from $\overrightarrow{x}_{1}$ to %%@
$\overrightarrow{P}(t)$. So,

\begin{equation}
\overrightarrow{P}(t)= \overrightarrow{q}(w_0) + t\overrightarrow{n}
\end{equation}
where
\begin{equation}
\overrightarrow{n}=\overrightarrow{e}_{12}\wedge \overrightarrow{e}_{13}
\end{equation}
Solving

\begin{equation}
|\overrightarrow{x}_{1}-\overrightarrow{P}(t)| =|\overrightarrow{x}_{4}-\overrightarrow{P}(t)|
\end{equation}
with $t\in(-2R_{0},2R_{0})$, we finally obtain the coordinates of the center of the maximal sphere, %%@
$\overrightarrow{P}(t)$, and
its radius $R$, which is simply given by $|\overrightarrow{x}_{1}-\overrightarrow{P}(t)|$.

\begin{figure}
\includegraphics[width=\columnwidth]{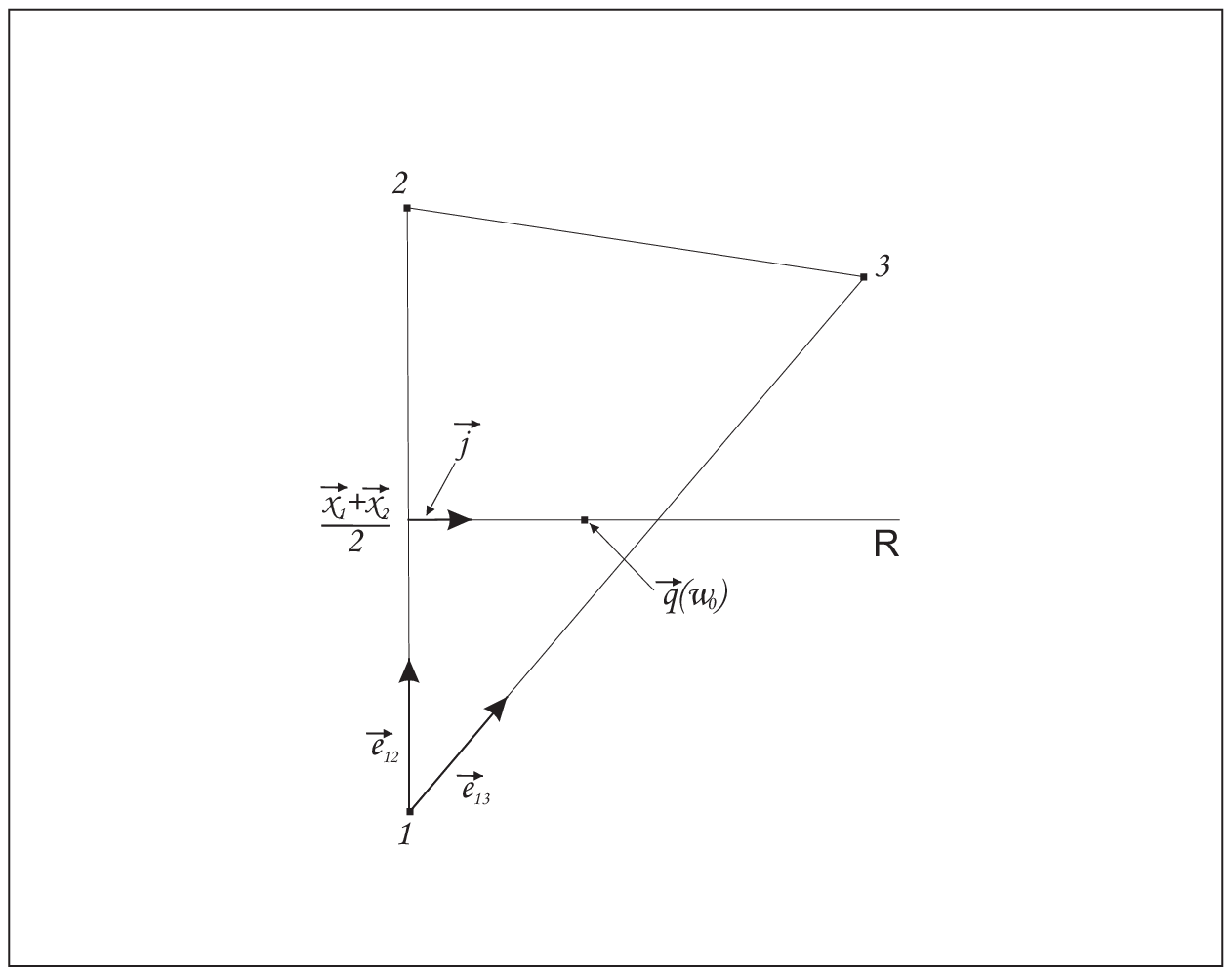}
\caption{}
\label{fig:A1}
\end{figure}


\begin{thebibliography}{}

\bibitem[Aikio \& M\"ah\"onen(1998)]{aikio} Aikio J., M\"ah\"onen P., 1998, \apj, 497, 534

\bibitem[Betancort-Rijo (1992)]{Juan92} Betancort-Rijo J.E., 1992, \pra, 45, 3447

\bibitem[Betancort-Rijo (1999)]{Juan99} Betancort-Rijo J.E., 1999, {\em J. Statistical Physics}, 98, 917

\bibitem[Betancort-Rijo (2005)]{Juan05} Betancort-Rijo J.E. et al., 2005, submitted to \apj, astro--ph/0509897

\bibitem[Colberg et al.(2005)]{colberg} Colberg, J.~M., Sheth, 
R.~K., Diaferio, A., Gao, L., \& Yoshida, N.\ 2005, \mnras, 360, 216 

\bibitem[Cole et al.(2005)]{cole05} Cole S., et al.~(the 2dFGRS team), 2005, \mnras, 362, 505

\bibitem[Cole et al.(1998)]{mockscole} Cole S., Hatton S., Weinberg D.H. \& Frenk C.S., 1998, \mnras, 300, 945

\bibitem[Colless et al.(2001)]{colless01} Colless M., et al.~(the 2dFGRS team), 2001, \mnras, 328, 1039

\bibitem[Colless et al.(2003)]{colless03} Colless M., et al.~(the 2dFGRS team), 2003, astro--ph/0306581

\bibitem[Conroy et al.(2005)]{charlie05} Conroy C., et al., 2005, submitted to \apj, astro--ph/0508250

\bibitem[Croton et al.(2004)]{croton} Croton, D.~J., et al.\ 
  2004, \mnras, 352, 828 

\bibitem[Croton et al.(2005)]{darren05} Croton, D.~J., et al.\ 
  2005, \mnras, 356, 1155 

\bibitem[Davis \& Peebles(1983)]{Davis1983} Davis, M., \& 
  Peebles, P.~J.~E.\ 1983, \apj, 267, 465
\bibitem[Einasto, Einasto, \& Gramann (1989)]{Einasto1989} Einasto J., Einasto M., Gramann M.,  1989, \mnras, 238, 155

\bibitem[Einasto et al.(1991)]{Einasto1991} Einasto J., Einasto M., Gramann M. \& Saar E., 1991, \mnras, 248, 593

\bibitem[Einasto et al.(1994)]{Einasto1994} Einasto J., Saar E., Einasto M., Freudling W. \& Gramann, M., 1994, \apj, %%@
429, 465

\bibitem[El-Ad H. \& Piran(1997)]{ElAd1997} El-Ad H., Piran T.,  1997, \apj, 491, 421

\bibitem[Fry(1986)]{Fry1986} Fry, J.~N.\ 1986, \apj, 306, 358 
 
\bibitem[Goldberg et al.(2005)]{Goldberg2005} Goldberg, D.~M., 
  Jones, T.~D., Hoyle, F., Rojas, R.~R., Vogeley, M.~S., \& Blanton, M.~R.\ 
  2005, \apj, 621, 643
\bibitem[Gottl\"ober et al.(2003)]{gottlober} Gottl\"ober S., Lokas E.,
Klypin A., Hoffman Y., 2003, \mnras, 344, 715

\bibitem[Hoyle \& Vogeley(2002)]{HoyleVogeley} Hoyle F., Vogeley M.S.,  2002, \apj, 566, 641

\bibitem[Hoyle \& Vogeley(2004)]{hoyle} Hoyle F., Vogeley M. S., 2004, ApJ, 607, 751

\bibitem[Hoyle et al.(2005)]{Hoyle2005} Hoyle, F., Rojas, R.~R., 
  Vogeley, M.~S., \& Brinkmann, J.\ 2005, \apj, 620, 618 
\bibitem[Kauffmann \& Fairall(1991)]{kauff} Kauffmann G., Fairall A.P., 1991, \mnras, 248, 313

\bibitem[Kirshner et al.(1981)]{Kirshner81} Kirshner R.P., Oemler A., Schechter P.L., Shectman S.A.,  1981,
  \apjl, 248, L57

\bibitem[Klypin \& Holtzman(1997)]{klypin} Klypin A. \& Holtzman J., 1997, astro--ph/9712217

\bibitem[Klypin et al.(1999)]{KlypinSat1999} Klypin A., Kravtsov A.V.,
  Valenzuela O., Prada F., 1999, \apj, 522, 82

\bibitem[Kravtsov, Klypin \& Khokhlov(1997)]{Kravtsov1997} Kravtsov A.V.,  Klypin A.A., Khokhlov A.M.,  1997, \apjs, %%@
111, 73

\bibitem[Maddox et al.(1990)]{maddox} Maddox S.J., Efstathiou G., Sutherland W.J. \& Loveday J., 1990, \mnras, 243, 692

\bibitem[Mathis \& White(2002)]{Mathis2002} Mathis H., White S.D.M.,  2002, \mnras, 337, 1193

\bibitem[Norberg et al.(2002)]{nor02} Norberg P., et al~(the 2dFGRS team) 2002, \mnras, 336, 907

\bibitem[Otto et al.(1986)]{otto} Otto, S., Politzer, D.H., Preskill J. \& Wise, M.B., 1986, \apj, 304, 62

\bibitem[Patiri, Betancort-Rijo \& Prada(2004)]{patiri} Patiri
S. G., Betancort-Rijo J. E., Prada F., astrop--ph/0407513

\bibitem[Peebles (2001)]{Peebles2001} Peebles P.J.E., 2001, \apj, 557, 495

\bibitem[Peebles(1980)]{Peebles1980} Peebles, P.~J.~E.\ 1980, 
 {\em The Large Scale Structure of the Universe}, 
Princeton University Press, 1980.

\bibitem[Plionis \& Basilakos (2002)]{Manolis} Plionis M. \& Basilakos S. \ 2002, \mnras, 330, 399

\bibitem[Rojas et al.(2005)]{Rojas2005} Rojas, R.~R., Vogeley, 
  M.~S., Hoyle, F., \& Brinkmann, J.\ 2005, \apj, 624, 571 
\bibitem[Rood(1981)]{rood} Rood H.J., 1981, Rep.Prog.Phys., 44, 1077

\bibitem[Rowan-Robinson(1968)]{RR} Rowan-Robinson M., 1968, \mnras, 138, 445

\bibitem[Sheth \& Van de Weygaert(2004)]{sheth} Sheth R.\ K.,
Van de Weygaert R., 2004, \mnras, 350, 517

\bibitem[Solevi et al.(2005)]{Solevi} Solevi P, Mainini R., Bonometto
S.A., Maccio' A.V., Klypin A. \& Gottloeber S., 2005, astro--ph/0504124

\bibitem[Tegmark et al.(2004)]{Tegmark} Tegmark, M., et al.\ 
2004, \apj, 606, 702
\bibitem[Springel, Yoshida \& White(2001)]{Gadget} Springel V.,
Yoshida N. \& White S.D.M., 2001, {\em New Astron.}, 6, 79

\bibitem[Springel et al.(2005)]{Springel} Springel, V. et al. 2005, {\emph Nature}, 435, 629

\bibitem[Van de Weygaert \& Van Kampen(1993)]{rien} Van de Weygaert R., Van Kampen E., 1993, \mnras, 263, 481

\bibitem[White(1979)]{White79} White, S.D.M., 1979, \mnras, 186, 145

\bibitem[York et al.(2000)]{york} York, D. G. et al. 2000, \aj, 120, 1579

\bibitem[Zehavi et al.(2002)]{Zehavi2002} Zehavi, I., et al.\ 
2002, \apj, 571, 172 

\end{thebibliography}
\end{document}